# Static and Dynamic Properties of Block-Copolymer Based Grafted Nanoparticles Across the Non-Ergodicity Transition

Daniele Parisi,[1,*] José Ruiz-Franco,[2] Yingbo Ruan (阮英波),[3,4] Chen Yiang Liu (刘琛阳),[3,4] Benoit Loppinet,[1] Emanuela Zaccarelli,[2,5] and Dimitris Vlassopoulos[1,6]

[1] Institute of Electronic Structure & Laser, FORTH, Heraklion 71110, Crete, Greece

[2] Dipartimento di Fisica, Sapienza Università di Roma, P.le A. Moro 5, 00185 Roma, Italy

[3] Beijing National Laboratory for Molecular Sciences, CAS Key Laboratory of Engineering Plastics, Institute of Chemistry, The Chinese Academy of Sciences, Beijing 100190, China

[4] University of Chinese Academy of Sciences, Beijing 100049, China

[5] CNR Institute of Complex Systems (ISC), Uos Sapienza, Roma, Italy

[6] Department of Materials Science & Technology, University of Crete, Heraklion 71003, Crete, Greece

[*] Corresponding author: daniele.parisi@iesl.forth.gr

## Abstract

We present a systematic investigation of static and dynamic properties of block copolymer micelles with crosslinked cores, representing model polymer-grafted nanoparticles, over a wide concentration range from dilute regime to an arrested (crystalline) state, by means of light and neutron scattering, complemented by linear viscoelasticity. We have followed the evolution of their scattering intensity and diffusion dynamics throughout the non-ergodicity transition and the observed results have been contrasted against appropriately coarse-grained Langevin Dynamics simulations. These stable model soft particles of the core-shell type are situated between ultrasoft stars and hard spheres, and the well-known star pair interaction potential is not appropriate to describe them. Instead, we have found that an effective brush interaction potential provides very satisfactory agreement between experiments and simulations, offering insights into the interplay of softness and dynamics in spherical colloidal suspensions.

# Introduction

Soft colloidal particles can be thought of as hybrids interpolating between polymers and hard spheres (HSs),[1] offering a plethora of possibilities for designing systems with tunable dynamic response. Examples of spherical soft colloidal systems are vesicles, dendrimers,[2] microgels,[3–5] block copolymer micelles,[6–9] polymer-grafted nanoparticles (PGNPs),[10–14] and star polymers.[15–19]

Unlike hard-spheres, for which the phase diagram and associated dynamic properties have been exhaustively investigated,[20] the respective consequences of softness have not been fully explored. For instance, bridging the gap between ultrasoft stars and polymer-grafted nanoparticles by identifying similarities and distinct features in the behavior still presents formidable challenges. The pioneering work of McConnnel et al.[21] established the phase diagram of polystyrene-polyisoprene diblock copolymers by means of small angle X-ray diffraction studies. Transition between face-centered cubic (fcc) and body-centered cubic (bcc) lattices was observed as the ratio of corona layer thickness to core radius increased. The resulting stable crystalline structure was found to be strongly dependent on the length scale of interactions: short-range repulsion favors fcc lattice over bcc and vice versa, as it was also demonstrated through previous experiments[22] and Monte Carlo simulations.[23] Particular attention has been paid on star polymers, whose density profile and conformation are described by the Daoud-Cotton model.[24–26] Such ultrasoft colloids represent one of the widest investigated model systems, where the softness can be tuned by varying the number of polymer chains anchored to a common center (functionality) and described accurately by a coarse-grained repulsive pair potential that varies logarithmically with the core-core distance.[27] Likos and co-workers[28] established the phase diagram of star polymer solutions in athermal solvent over a wide range of number densities and functionalities. The observed phases included bcc, fcc, body centered orthogonal (bco) and diamond crystals. Over a limited range of number of arms, a reentrant melting of the bcc phase was observed as the packing fraction increased. Additionally, at relatively high functionality a fcc phase was detected instead, and the effect of increasing packing fraction translated into different ordered lattices until the diamond crystal was reached. Interestingly, this interaction potential was found to be suitable also for several star-like systems, such as diblock copolymer micelles.[8,29–31] The common denominator of such systems seems to be the relatively large ratio of hydrodynamic radius of the particle to the core radius (between 7 and 10). Gupta et al.[32] have presented a thorough experimental and theoretical phase diagram of soft colloids made of tunable block copolymers which allowed adjusting the softness by altering the solvophobic-to-solvophilic block ratio, hence bridging the star-to-micelle regimes. An important finding relates to the threshold in functionality above which crystalline phase is not formed and the resulting non-ergodic state is characterized by a random distribution of the particles forming a glass. On the other hand, below the threshold value, fcc lattice or fcc-to-bcc transition is observed as the previous works mentioned above. Note that these micelles were stable due to the combination of solvophobicity and large interfacial tension. In general, block copolymer micelles are characterized by arm exchange kinetics which impacts their structure and dynamics properties.[8,30,33–39] Stable micelles can be obtained in a relatively straightforward way by chemical crosslinking of their cores, and this type of soft colloids will be discussed below.[5,10,40,41]

An important challenge is the fact that the internal particle microstructure may give rise to completely different features with respect to the star-polymers or star-like micelles, in a way that is not fully understood yet. For instance, Ohno *et al*.[42] and Morinaga *et al*.[43,44] showed by means of confocal laser scanning microscopy in fluorescent mode, that for core-shell systems (silica core grafted with polymethylmethacrylate) the crystalline phase was strongly dependent on the brush length but the bcc crystalline structure was not observed at all. Instead, a hexagonal close-packed (hcp) structure was detected. When short chains were grafted, the probability of having a fcc lattice was found to be the same as for the hcp organization. A nearly random stacking of fcc and hcp was formed. The authors called this region concentrated polymer brush (CPB) where the excluded volume region effect is irrelevant. In a region of large chain lengths, the CPB was followed by the excluded volume region, or semidilute polymer brush regime (SDPB), whose effect on the crystalline structure becomes important. Note that, the brush conformation regime proposed by Ohno *et al*.[43] derives from the well-known Daoud-Cotton model for star polymers.[24] The probability of finding a fcc increases sharply as the interaction potential softens due to the increase in length of the brushes. However, during the crystallization process, some hcp arrangements might be frozen in an irreversible way. This non-equilibrium process is more likely to happen for systems with a shorter range of interparticle potential such as hard-spheres and CPB systems. The theoretically deduced entropy difference between fcc and hcp phases may be too small to realize an equilibrium system experimentally.[43] On the other hand, as the interparticle potential range becomes longer-ranged, not only the nearest-neighbor but also the second-nearest (and higher-order) interactions can play a role in crystallization, yielding much larger energetic and entropic differences between fcc and hcp phases. Therefore, crystalline processes are less likely to occur in systems with a longer range of interparticle potential.

Given their static features, soft colloids also exhibit very rich dynamics, and numerous experimental works have revealed intriguing properties.[40,45–52] Semenov *et al*.[48] investigated the dynamics of polybutadiene multiarm star polymers in good solvency conditions by means of dynamic light scattering. Three relaxation modes in the semidilute regime were probed: a fast cooperative diffusion due to the polymeric nature of the systems (star-arms interpenetration taking place at the overlap concentration c* and beyond), a slow self-diffusion of the star-polymers (due to finite functionality, hence size polydispersity) and an intermediate structural mode reflecting the collective structural rearrangements of ordered stars. With increasing concentration, the fast-cooperative diffusion speeds up,[53–55] whereas the slowest mode slows-down, while the collective structural mode is concentration-independent. Along the same lines, Loppinet et al.[40] investigated concentrated suspensions of diblock copolymer micelles with crosslinked cores and also detected three relaxation modes analogous to the star polymer case. The study of the dynamics of core-shell (silica-polystyrene) particles by Voudouris et al.,[50] revealed similar behavior. However, by varying the molecular architecture (grafting density, aggregation number, and molar mass of the grafted chains) it was found that high-grafting density particles exert a large osmotic pressure which makes the grafted chain interpenetration more difficult, shifting the emergence of a polymeric cooperative diffusion to concentrations well above c*. In addition, it was shown that the decoupling of the modes becomes more visible for systems with low grafting density and long brushes.[50]

From the above brief summary, it is evident that the slowest mode, i.e., the self-diffusion of the particle, can be observed in hard-spheres, multiarm star-polymers and core-shell particles. On the other hand, the cooperative diffusion and arm retraction modes represent unique signature of interpenetration of hairy particles. In this respect, we note that microgels without dangling ends do not exhibit cooperative diffusion relaxation mode.[56]

In spite of the numerous similarities in phase diagram and dynamics among soft spherical colloids, their realm remains quite broad and there is a need to link soft interactions to structure and dynamics over a wide range from stars to grafted spheres.[57] Clearly, this cannot be done with one type of interaction potential, as we show here. We thus need further experiments with model systems and appropriate description of their interactions.

In this work we use well-characterized polymer-grafted nanoparticles (PGNPs), which are stable (since their cores are crosslinked) and exhibit characteristics embracing core-shell particles and star-polymers. Suspensions of PGNPs are investigated by means of, neutron and light scattering, as well as rheology in the linear viscoelastic regime. Langevin Dynamics (LD) simulations, using the star polymer interaction potential, are performed and the results compared with the observed static and dynamic properties of the PGNPs. While the static structure factor can be captured by the star polymer potential, major issues arise when the same potential is used to test the dynamic behavior. In fact, as the concentration increases, the star-polymer potential is not able to capture the experimental intermediate scattering functions. To this end, the pair potential developed in the context of colloidal interfaces with adsorbed gelatin[58] (hereafter named "brush potential") is modified and applied to our PGNPs. It is found to be very effective in capturing both static and dynamic properties observed in experiments conducted at various concentrations and length scales. In addition to the above-mentioned finding, it is also reported that: i) an arrested state is detected by means of static light scattering and rheology, ii) the solid-like behavior has features of a crystalline phase and given the polymer chain being in the SDPB regime, a fcc structure is expected according to Morinaga *et al.*[44] Indeed, LD simulations confirm the presence of a fcc structure. iii) Dynamic light scattering at various concentrations revealed three different modes assigned to the cooperative diffusion of the polymer chains, center of mass motion (collective diffusion) and self-diffusion. Our results show the importance to carefully combine different approaches to study these systems, the richness of softness and its link to structure and dynamic properties. The present PGNPs serve as another model soft colloidal systems, easier to prepare compared to star polymers, stable and spanning the regime from stars toward hard spheres.

## Materials and Methods

Spherical grafted nanoparticles were prepared by microphase separation of diblock copolymers of poly(3-(triethoxysilyl)propyl methacrylate) (PTEPM) which represents the core, and polystyrene (PS) the shell, in the presence of oligomers of both o-TEPM and/or o-S, followed by cross-linking and dispersion in good solvent for polystyrene. This technique, known as Assembly Cross-linking and Dispersion (ACD), allows to obtain precisely tailored nanoparticles which include different shapes with identical PS shells, different core sizes but the same shell, and particles with fixed shape but varied PS shell. Thus, among the main advantages of this technique there is the possibility to obtain nanoparticles with different shape, same shell and very similar grafting density. Further details about these nanoparticles are reported elsewhere.[10,59]

Table 1 reports the molecular characteristics of the spherical micelles used in the present work; core radius ($R_c$), polydispersity index (PD) of the polystyrene chains, number-average molar mass of the PS arm ($M_{n,arm}$), PS weight fraction respect to the total particle (*PS fraction*), grafting density ($\rho$), aggregation number ($N_{agg}$), and the weight-average molar mass ($M_w$) of the whole nanoparticle calculated as $M_w = (M_{n,arm} \cdot PDI)/PS\ fraction$.

Two solvents, toluene and chloroform were used for the scattering experiments. The hydrodynamic radius ($R_H$), the radius of gyration ($R_g$), their ratio ($R_g/R_H$) and the overlap concentration (c$^*$) for each solvent were measured by means of light scattering (see Figure S1 in the SI) and are reported in Table 2. Concentrations are expressed in terms of an effective volume fraction ϕ (simply called volume fraction hereafter) estimated as the ratio between the actual and the overlap concentration defined as $c^* = \frac{M_w}{\frac{4}{3}\pi R_H^3 \mathcal{N}_A}$ with $\mathcal{N}_A$ being the Avogadro number. Given the soft hairy nature of the particles, the effective volume fraction may reach values higher than the unit values due to its deformability and compressibility.

Table 1. Molecular Characteristics of the PGNPs.

| $R_c^a$ [nm] | $PDI^a$ [-] | $M_{n,arm}^b$ [kg/mol] | PS fraction [wt%] | $\rho^c$ [nm$^{-2}$] | $N_{agg}^d$ [-] | $M_w$ [kg/mol] |
|---|---|---|---|---|---|---|
| 15 | 1.04 | 111 | 80 | 0.075 | 212 | 3·10$^4$ |

[a] Core size and polydispersity index estimated through Transmission Electron Microscopy (TEM) images in dry state (see Figure S2 in the SI). [b] Molar mass of the tethered PS chains. [c] Grafting density. [d] Number of grafted PS chains. See Y. Ruan *et al*.[10]

Table 2. Hydrodynamic size, radius of gyration and overlap concentration.

| Solvent | $R_H$ [nm] | $R_g$ [nm] | $R_g/R_H$ [-] | $c^{*a}$ [g/ml] |
|---|---|---|---|---|
| *Toluene* | 62 | 58 | 0.94 | 0.05 |
| *Chloroform* | 68 | 53 | 0.78 | 0.04 |

[a] Overlap concentration $c^* = 3M_w/(4\pi R_H^3 N_A)$

The ratio between the radius of gyration and the hydrodynamic radius is an indication of the softness of the colloidal system.[60] Such characteristic particle sizes strongly depend on solvency conditions.[61] In the present study, $R_g/R_H$ decreases as the solvent is varied from toluene to chloroform. Notably, the $R_g/R_H$ ratio ranges from 0.77, for a homogeneous sphere such as a hard sphere, to 1.5 for monodisperse linear chains in good solvent.[60] In the present systems, as well as in core shell particles,[50] star-like[40,45] and star polymers,[62] such a ratio is > 0.77. However, recent investigations on thermoresponsive ionic microgel systems showed a $R_g/R_H$ ratio < 0.77.[63] In our study, we find that the lowest value belongs to the solution in chloroform, where nanoparticles nearly approach the hard-sphere limit. However, in the latter case the $R_H$ resulted in the highest value, more than that in toluene in spite of the fact that toluene is a good solvent for PS. This can be tentatively ascribed to the fact that chloroform favors better solvency conditions for the crosslinked core promoting its slight swelling. The procedure adopted in order to estimate the hydrodynamic radius and the radius of gyration is reported in Figure S1 of the SI. Note that, for the present systems, the ratio of overall hydrodynamic-to-core radii is about 5, smaller than the soft systems previously mentioned in the Introduction. The refractive indices of the core, the shell, and the different solvents used in this work are reported in Table 3.[64]

Table 3. Refractive index of materials at λ = 532 nm.[64]

| Material | Refractive index at $\lambda$ = 532 nm |
|---|---|
| *Polystyrene* | 1.5983 |
| *Poly(3-triethoxysilyl propyl methacrylate)* | 1.460 |
| *Toluene* | 1.5019 |
| *Chloroform* | 1.4471 |

The interaction between PGNPs are purely repulsive. Static light scattering measurements in the dilute regime in toluene allowed the measure of the second virial coefficient, $A_2$, that characterizes the particle-particle interaction. The Zimm representation[65] is presented in Figure S3 in the SI, together with the linear fit which yielded a value $A_2$ =3.4 $10^{-5}$ mol ml g$^{-2}$. The second virial coefficient $A_2$ resulted positive, implying good solvency conditions, but its value is almost one order of magnitude lower than that of pure PS chains in toluene at the same $M_w$, 1.1 $10^{-4}$ mol ml g$^{-2}$.[66] As expected, particle interactions are different from those between simple linear chains. This relates to the colloidal nature of the PGNPs.

## Photon Correlation Spectroscopy

The experimental normalized light scattering intensity $I(q,t)$ used to compute the autocorrelation function[67] $G(q,t) \equiv \langle I(q,t)I(q)\rangle/|\langle I(q)\rangle|^2$ was measured over a broad time range ($10^{-7}$–$10^3$ s) with an ALV-5000 goniometer/correlator (ALV, Germany) setup using an Nd:YAG laser (Oxxius, France) at λ=532 nm. The scattering wavevector, defined as $q = (4\pi n/\lambda)\sin(\theta/2)$, where n is the refractive index of the medium and θ the scattering angle, was varied in the range between 0.005 and 0.035 nm$^{-1}$. Under homodyne beating conditions, the desired intermediate scattering function is computed from the experimental $G(q,t)$: $C(q,t) = [G(q,t)-1]/f^*]^{1/2}$ where *f*\* is an instrumental coherence factor which is typical smaller than one. In the dilute regime $C(q,t)$ is a single decay function, and the effective short time diffusion coefficient is determined from the initial decay rate according to[68]

$$D_{Sh} = \frac{\Gamma}{q^2} = \left(\frac{1}{q^2}\right)\lim_{t=0}\left(\frac{d[\ln g^{(1)}(q,t)]}{dt}\right) \qquad (1)$$

where $g^{(1)}(q,t)$ is the normalized intermediate scattering function $C(q,t)$. In the non-dilute regime, the presence of multiple relaxation processes requires the analysis by inverse Laplace transformation:

$$C(q.t) = \int L(\ln \tau) \exp(-t/\tau) \, d\ln \tau \qquad (2)$$

The distribution of relaxation times $L(\ln \tau)$ can be decomposed as a sum of distributions $L(\ln \tau) = \sum_i L_i(\ln \tau)$. The intensity of the *i*-th process can be obtained from the area of the corresponding distribution $L_i(\ln \tau)$, $I_i(q) = I(q) \int_{\ln \tau} L_i(\ln \tau) d \ln \tau$, with $I(q)$ being the total scattering intensity. The relaxation rate $\Gamma_i$ is obtained from the peak position of $L_i(\ln \tau)$. The analysis of the $L(\ln \tau)$ yields the diffusion constants (for the diffusive processes), the rate of relaxation as well as the associated intensities $I_i$, which provide a measure of the efficiency of the particular motional mechanism to relax the density fluctuations of the system.[50,69]

The intermediate scattering functions C(q,t) were analyzed either with a double Kohlrausch-Williams-Watts (KWW) exponential equation with the form $C(t) = A_1 e^{-(\frac{t}{\tau_1})^{\beta_1}} + A_2 e^{-(\frac{t}{\tau_2})^{\beta_2}}$, with *t* being the time, $\beta_1$ and $\beta_2$ the stretching exponents, $\tau_1$ and $\tau_2$ characteristic relaxation times of the system, and $A_1$ and $A_2$ constants, or with the CONTIN algorithm.[70]

Dense suspensions of hairy particles are well known to give rise to complex intermediate scattering functions. The analysis of such correlation functions is challenging especially in the view of the multiple solutions available that will provide an equivalent fit of the experimental functions with different physical interpretation (see Figures S4 and S5 in the SI). Here we adopted a three-mode analysis that has been used for other soft colloidal particles.[40,45,48,50] This can be understood as a simple visualization of the dense dispersion of cores in a sea of polymeric blobs.[48] Such a simple model implies the existence of three relaxation modes. The fastest mode relates to the shell polymer versus solvent motion. It is usually referred as the cooperative mode in semi-dilute polymer literature.[48,54] The two other modes relates to the core diffusion. The intermediate mode relates to the collective diffusion of the cores. Owing the inherent "scattering polydispersity" of the type of systems, a third mode is expected and relates to the "incoherent" self-diffusion of the core.

## Neutron scattering

Small Angle Neutron Scattering (SANS) experiments were performed at Laboratoire Léon Brillouin in Saclay (France). This technique was used for solutions of spheres in perdeuterated toluene at four different volume fractions, 0.034, 0.2, 0.5, 1.6, ranging from dilute to concentrated regime. Hellma quartz cuvettes (with thickness of 1 mm) were used for the experiments. Neutrons were used with a wavelength of $\lambda = 0.63$ nm and $\frac{\Delta\lambda}{\lambda} \approx 0.18$, at three detector distances (2, 8, and 20 m) in order to access the scattering wavevector range $0.02 < q < 1.5$ nm$^{-1}$. The scattered neutrons were detected with a two-dimensional $^3$He detector consisting of 64 channels, each having a width of 0.8 cm. The count rate was controlled in a way that minimized the dead time effects. The isotropic raw data were radially averaged, corrected for background scattering, and the intensity was converted into absolute units by using a polyethylene sample of known incoherent scattering as standard (calibrated against vanadium).

## Linear viscoelasticity

Small amplitude oscillatory shear experiments were performed with a sensitive strain-controlled ARES (TA, USA) rheometer with a force balance transducer 100FRTN1. The temperature was controlled by means of a Peltier element with an accuracy of ± 0.1 °C connected to a recirculating water/ethylene glycol bath. A cone-and-plate configuration was used, where the cone was a 8 mm diameter home-made stainless steel with angle 0.166 rad and truncation gap 210 μm, whereas the plate was the Peltier unit itself. Rheological experiments were performed only in toluene at 25 °C and the measuring area was also sealed with a home-made solvent trap containing the same solvent to saturate the environment and reduce the risk for evaporation. Note that no rejuvenation-aging protocol was applied. The reason lies in the fact of minimizing the measuring time due to the high volatility of the solvent. Indeed, the scope of such an experiment was only to validate the rheological state of the suspension and corroborate the ergodic-to-nonergodic transition observed with light scattering.

## Langevin Dynamics (LD) simulations

The properties of hairy particles are usually analyzed in the context of the star polymer potential developed by Likos.[27] The latter represents a coarse-grained potential for long-ranged repulsive interactions which depends on the functionality. For spherical particles with adsorbed polymers forming an effective brush, Likos *et al.*[58] developed a pair potential specifically accounting for the interactions between colloidal surfaces with adsorbed gelatin. This was accomplished through a naïve superposition of a simple electrostatic repulsion with an equally simple model for polymer steric stabilization. By only adopting the term that considers the steric repulsion, in the specific case given by the polymeric shell of the soft particle, it is possible to reproduce with good agreement the static and dynamic properties of the systems considered in this work. A similar approach was also adopted by Loppinet *et al.* for stable diblock copolymer micelles with non-negligible core and ratio of hydrodynamic particle to core equal to 5.5.[45] The star polymer potential $V_{SP}(r)$ is

$$\beta V_{SP}(r) = \begin{cases} \dfrac{5}{18} f^{3/2} \left[ -\ln\left(\dfrac{r}{\sigma_{SP}}\right) + \left(1 + \dfrac{\sqrt{f}}{2}\right)^{-1} \right] & r \leq \sigma_{SP} \\ \dfrac{5}{18} f^{3/2} \left(1 + \dfrac{\sqrt{f}}{2}\right)^{-1} \left(\dfrac{\sigma_{SP}}{r}\right) \exp\dfrac{-\sqrt{f}(r - \sigma_{SP})}{2\sigma_{SP}} & r > \sigma_{SP} \end{cases} \quad (3)$$

where $f$ is the functionality and $\sigma_{SP}$ is a characteristic length which, for large $f$, is found to be comparable to the hydrodynamic radius of the particle.[8,71,72]

On the other hand, the brush potential $V_{Brush}(r)$ reads

$$\beta V_{Brush}(r) = \begin{cases} \infty & r < 2R_c \\ f(y) & 2R_c < r < 2(R_c + L) \\ 0 & 2(R_c + L) < r \end{cases} \quad (4)$$

where $y = (r - 2R_c)/(2L)$ and $f(y)$

$$f(y) = \dfrac{16\pi R_c L^2}{35 s^3} \left[ 28(y^{-1/4} - 1) + \dfrac{20}{11}(1 - y^{11/4}) + (y - 1) \right] \quad (5)$$

with $R_c$ the radius of the core, $L$ the thickness of the brush layer and *s* the square-root of the inverse of the grafting density defined as $s = \sqrt{\dfrac{A}{N_{agg}}} = 3.65$, with A being the core surface and $N_{agg}$ the aggregation number reported in Table 1. In both potentials, β represents the inverse thermal energy $1/k_B T$, with $k_B$ being the Boltzmann constant and T the temperature. The potential in Eq.4, for distances smaller than $2R_c$, resembles that of hard-spheres, whereas the interpenetration of the brush layer for $2R_c < r < 2(R_c + L)$ is modelled by a smooth function *f(y)*. For such a potential we define the total particle size $2(R_c + L) = \sigma_{Brush}$. Note the different definition of the characteristic length in the two potentials: for the brush model, $\sigma_{Brush}$ corresponds to the diameter of the particle, whereas for the star polymer potential $\sigma_{SP}$ corresponds to an effective radius, amounting to 1.3$R_g$, or similarly, to the hydrodynamic radius.[8,71,72]

Figure 1 depicts a comparison between star and brush potentials at the same number of arms ($f$ = 212), highlighting their qualitative difference and the smoother divergence of the former potential as particle-particle distance is reduced. The brush potential appears to be closer to that of hard-spheres, so it is more relevant to grafted spheres with substantial core (as discussed below). Note that for the brush potential that we consider in this work we fix the parameters to experimental values, namely the radius of the core ($R_c$) and the square-root of the grafting density ($1/s$). In view of this, we only adjust the brush thickness (L), and consequently the particle size ($\sigma_{Brush}$), and the packing fraction ($\phi_{Brush}$). Remarkably, we find that the brush potential provides more physical insights on the architectural characteristics of the experimental systems.

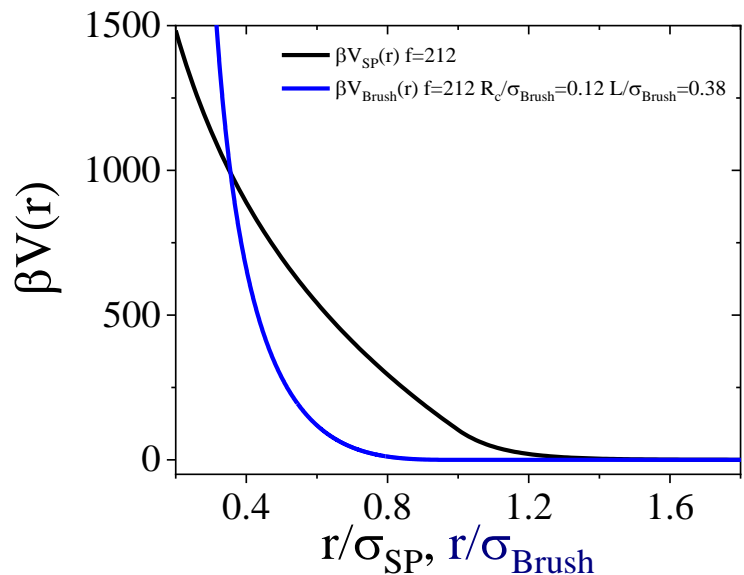

Figure 1. Comparison between star polymer and brush interaction potentials at the same number of arms, $f = N_{agg} = 212$. In addition, the quantities $R_c/\sigma_{Brush} = 0.11$, $L/\sigma_{Brush} = 0.39$ $s = \sqrt{\frac{A}{N_{agg}}} = 3.65$ of the brush potential correspond to those of the real system. Importantly, whereas $\sigma_{Brush}$ is the total particle diameter, $\sigma_{SP}$ is a characteristic length which corresponds to the hydrodynamic radius of the star polymer.[8,71,72]

We performed Langevin Dynamics simulations of a system with N = 2000 particles with mass $m$ = 1, where the total force on the $i$th particle is defined as:

$$F_i = F_i^C + F_i^D + F_i^R \qquad (6)$$

where $F_i^C$ represents the conservative force computed from the star polymer and brush interaction potentials. $F_i^D$, defined as $F_i^D = -\xi v_i$, where $\xi$ is the friction coefficient, fixed in our case at $\xi = 10^2$, and $v_i$ is the particle velocity. Random forces $F_i^R$ are characterized by $<F_i^R> = 0$, and $<F_i^R(t)F_j^R(t')> = 6\xi k_B T \delta(t-t')$. The equation of motion were integrated using the self-adaptive OVRVO scheme,[73] which is suitable for both equilibrium and nonequilibrium dynamics.[73,74] A simulation time-step $dt = 0.002$ was used. Mass, length and energy are measured in units of m, σ and $k_B T$, whereas the simulation time is in units of $t = \sqrt{(m\sigma^2)/(k_B T)}$. The radius cutoff is $r_c = 1.5\sigma_{Brush}$ for the brush potential, whereas $r_c = 3\sigma_{SP}$ for star potential. Since both potentials are athermal and temperature is not varied in experiments, the value of the temperature is kept constant, i.e. $k_B T = 1$. To study the dynamic properties of the grafted nanoparticles, a polydispersity of 10% was used to avoid crystallization at high densities.[75] In addition, we also performed selected simulations of monodisperse particles interacting with the brush potential to investigate the possible crystalline phase displayed by such soft systems.

The effective packing fraction is defined as $\phi_{Brush} = \frac{\pi}{6}\sigma_{Brush}^3 \frac{N}{L_{Box}^3}$ and $\phi_{SP} = \frac{\pi}{6}\sigma_{SP}^3 \frac{N}{L_{Box}^3}$ for the brush and star potentials, respectively, where N is the number of particles and $L_{Box}$ the size of the simulation box. The latter parameter, and hence the packing fraction, was varied until the numerical results qualitatively match the experimental observations (see SI for further details). Note that, since the particle size is defined in a different way in the two potentials, different values of the packing fraction can be obtained. However, the concentration regime probed is the same as the experimental one.

## Results and discussion

## Structural properties

All experiments reported in this section were performed in toluene. The effect of concentration was investigated through static light scattering measurements and the intensity, expressed in terms of $R_\theta/KC$, is plotted against the scattering wavevector q, as shown in Figure 2A. The smallest length scale that can be probed within the explored range of q is of the order of the particle size (~ 60 nm).

Three regimes can be identified: I) a dilute (non-interacting) regime when $0.045 < \phi < 0.3$ with weak q-dependence of the intensity. II) In the semi-dilute regime, $0.3 < \phi < 1.26$, a strong q-dependence of the intensity is observed as a result of particle interactions and crowding. III) At even higher volume fractions ($\phi = 1.26$ and beyond), a further drop in the intensity occurs which reflects the presence of an arrested state. The latter represents a non-ergodic response, which in this particular case, displays all the characteristics of a fcc crystalline structure, as will be shown later. While in the present case the non-ergodicity is driven by particles crowding, i.e. increase in concentration, there are also reported cases of arrested states on polymer-grafted nanoparticles[76] and star polymers[77] mediated by the temperature. The transition to a solid-like behavior was also confirmed by small amplitude oscillatory shear experiments. To this end, Figure 2B depicts dynamic frequency spectra in toluene at two different volume fractions, $\phi = 0.4$ and $\phi = 1.4$, belonging to regime II and III, respectively. Indeed, between regimes II and III, a liquid-to-solid transition emerges. At $\phi = 1.4$, the storage modulus $G'$ is frequency-independent and larger than the loss modulus $G''$ in the whole frequency range probed in the experiments. The plateau value of the storage modulus reaches a value of about 250 Pa. Such a value is much lower than that of a typical colloidal glass of (other) soft particles investigated in the literature.[18,19,29,77–79] This simple observation is reminiscent of the behavior of hard sphere suspensions[80] which reveals that a colloidal crystal is characterized by lower dynamic moduli compared to the respective repulsive glass. Whereas this is a simple observation, it certainly motivates more investigations in the direction of linking the viscoelastic response of soft colloidal crystals and their respective glasses. Conversely, the $\phi = 0.4$ sample that belongs to regime II, exhibits the typical response of a viscoelastic liquid. Note that the frequency-dependent moduli of the viscoelastic liquid at $\phi = 0.4$ do not exhibit the expected 1 and 2 terminal slopes for a fully relaxed system. The reason can be tentatively ascribed to the lack of rejuvenation protocol applied, hence, a not fully homogeneous system. In fact, due to the high volatility of the solvent and despite the precautions taken, the experiments were performed immediately after the sample loading.

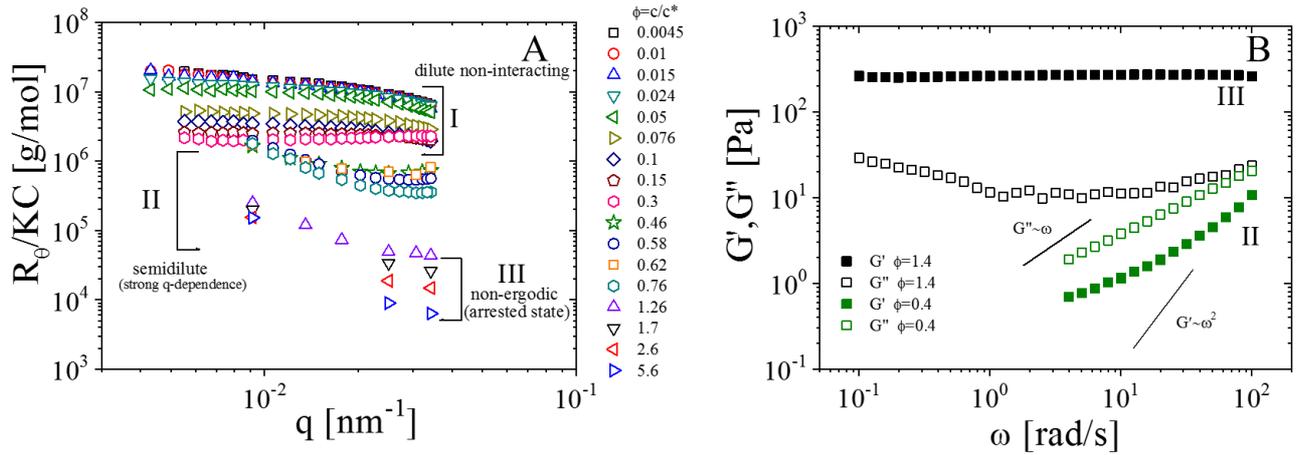

Figure 2. A) Total static light scattering intensity in terms of $R_\theta/KC$ versus the scattering wavevector q at various volume fractions. From the dilute non-interacting regime to an arrested state (crystal). There are three regimes: (I) dilute non-interacting at $\phi < 0.3$, (II) semi-dilute at $0.3 < \phi < 1.26$, with strong q-dependence, and (III) non-ergodic at $\phi > 1.26$, characterized by a fcc crystalline structure (read text). B) Storage modulus G′ (solid symbols) and loss modulus G″ (open symbols) as a function of the oscillation frequency ω at 0.01 shear strain units (well within the linear viscoelastic regime) for $\phi = 0.4$ (green symbols) and $\phi = 1.4$ (black symbols). All the experiments were performed at 25 °C.

The question whether the arrested non-ergodic state has the characteristics of a glass or an ordered structure arises naturally. Whereas for soft systems like star polymers, a glass transition has been readily identified slightly above the overlap concentration (depending on functionality and arm molar mass),[78,81,82] with crystalline phase being hard, albeit not impossible to detect,[83,84] core-shell particles like micelles have been reported to easily crystallize, as also mentioned in the Introduction. Indeed, fcc or bcc phases form depending on the degree of polymerization of the tethered chains,[44] provided that the polydispersity of the spherical particles is small. To this end, small-angle neutron scattering (SANS) experiments were performed on PGNP solutions in a higher range of q. The normalized intensity rescaled by the volume fraction is shown as function of q in Figure 3. Light scattering data are also rescaled by the volume fraction and plotted along with the SANS data in order to extend the q-range to lower values. At high q-values the intensity curves overlap and the intensity decay follows a power-law with an exponent of -2, which is reminiscent of Gaussian coils in good solvency conditions.[85] At $\phi = 0.034$, still in the dilute non-interacting regime, the particle form factor can be obtained and a good overlap with light scattering data is attained, although there is a slight difference in concentration (concentrations are reported in the legend of Figure 3). As the volume fraction approaches 0.5, structural peaks emerge. However, the solution is still in an ergodic state and the length scale at the peak (green triangles) is 2π/0.07 = 89 nm, slightly larger than the hydrodynamic size of the particle in dilute regime ($R_H$ = 62nm). A further increase in volume fraction leads to an arrested state and the structural peaks shift towards higher q-values. In fact, the peak now corresponds to 2π/0.12 = 52 nm < $R_H$, implying a strong packing of the particles (likely reflecting osmotic compression and interpenetration). A pseudo-structure factor was estimated by simply dividing the total scattering intensity I(q) at ϕ = 1.26 by the form factor P(q) at ϕ = 0.034 and plotted versus the scattering wavevector (Figure 4). This approach entails the rough approximation to P(q) being concentration-independent, which is not entirely true. Indeed, as the particles deform and reduce their size with concentration it is not appropriate to define such a ratio as the static structure factor of the system. Nevertheless, it turns to be very useful to implement an effective interaction potential for the LD simulations. The value of structural peak detected in Figure 4 overcomes the empirical Hansen-Verlet crystallization threshold value of 2.85,[86,87] suggesting the possible presence of an ordered state or a mixture of ordered states. As a matter of fact, it is not possible to distinguish different crystalline phases as shown by the fitting curves in Figure 4, since the structure factors for an fcc, bcc as well as a hcp order would all fit reasonably well the experimental data with weak and noisy higher-order peaks. The presence of crystalline phases in experimental soft systems is of course possible[7,21,44,88,89] but less likely in comparison to monodisperse hard spheres[20,80] and spherical micelles with dynamic arm exchange.[8,33,34] According to Ohno *et al.*,[43] the probability of finding an fcc structure increases as the degree of polymerization of the tethered chains increases. In fact, this occurs in the presence of an excluded volume region (semi-dilute polymer brush regime). By knowing the grafting density ρ at the core, the radius of the core $R_c$ (see Table 1) and the monomer length $l_m$ (0.25 nm[90]), a critical radius is estimated as $r_{crit} = R_c \rho^{1/2} l_m v^* = 0.7\ nm$ where $v^* = 4\pi^{1/2} v$ with $v$ being the excluded-volume parameter.[24] Assuming good solvency conditions, $v = 0.5$. Given that $R_c \gg r_{crit}$[43], one may infer that the entire brush layer is in the SDPB regime, hence, the most likely structure is represented by an fcc order.

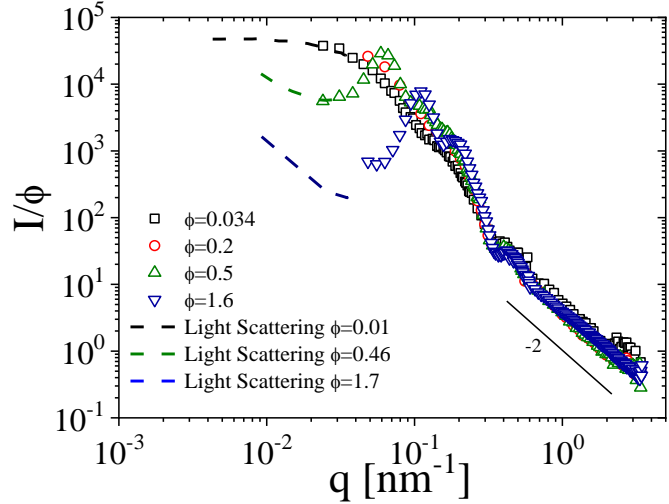

Figure 3. Neutron scattering experiments for four different volume fractions in toluene at 25 °C in terms of normalized intensity divided by the volume fraction against the scattering wavevector q. Dashed lines represent light scattering data at similar volume fraction (see also Figure 2). The slope -2 at high q values represents the q-dependence of the intensity for Gaussian chains in good solvent.

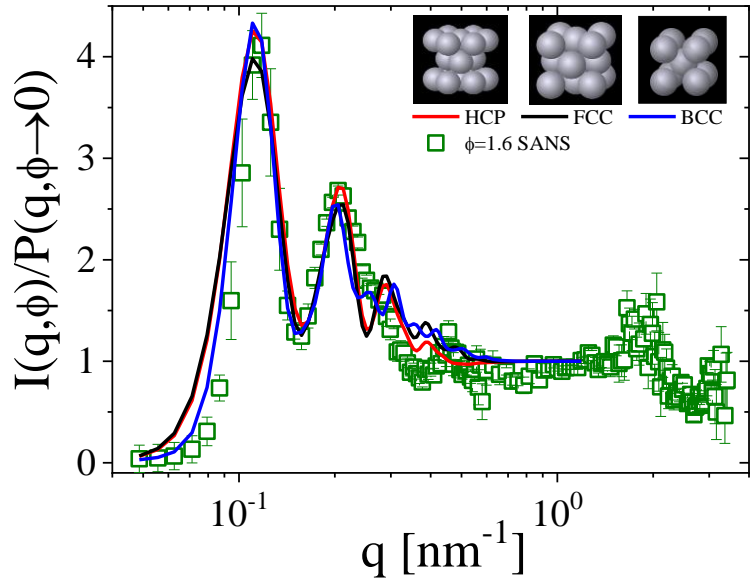

Figure 4. Pseudo-structure facture obtained by dividing the scattered intensity at $\phi = 1.6$ by the form factor at $\phi = 0.034$. Black, blue and red lines are the fcc, bcc and hcp crystalline structure factors, respectively. According to the Hansen-Verlet criterion the first peak overcomes the 2.85 threshold value for crystalline structure.[86]

We simulated monodisperse particle suspensions and performed the local bond-order analysis introduced by Steinhardt et al.[91]. The order parameters are expressed in terms of angular correlations between a vector, characterized by a certain number of spherical harmonics identifying the local environment around a central particle, and its neighboring vectors. This leads to the rotationally invariant order parameter distributions $P(\overline{w}_i)$, with $i$ being the number of spherical harmonics, which are used to discriminate between bcc, fcc and hcp structures.[92–94] The probability of finding either bcc or hcp/fcc structures is shown in Figure 5A in terms of $P(\overline{w}_6)$ for two different volume fractions corresponding to regime III. Figure 5B shows that, when considering 4 spherical harmonics, $P(\overline{w}_4)$, the most likely crystalline structure is a fcc, i.e., the same that was previously assessed, based on the brush conformation model proposed by Ohno *et al*.[43] Two important remarks follow in order. Simulations were performed by using the effective brush potential, for reasons that will be clarified later, and star polymers with the same functionality as the PGNPs would also exhibit a fcc crystalline structure.[95]

The system's tendency to crystallize is compatible with the formation of a fcc crystal. Additional studies would be needed to investigate the stable crystals in the full phase diagram of the brush potential. However, this goes beyond the scope of the present work.

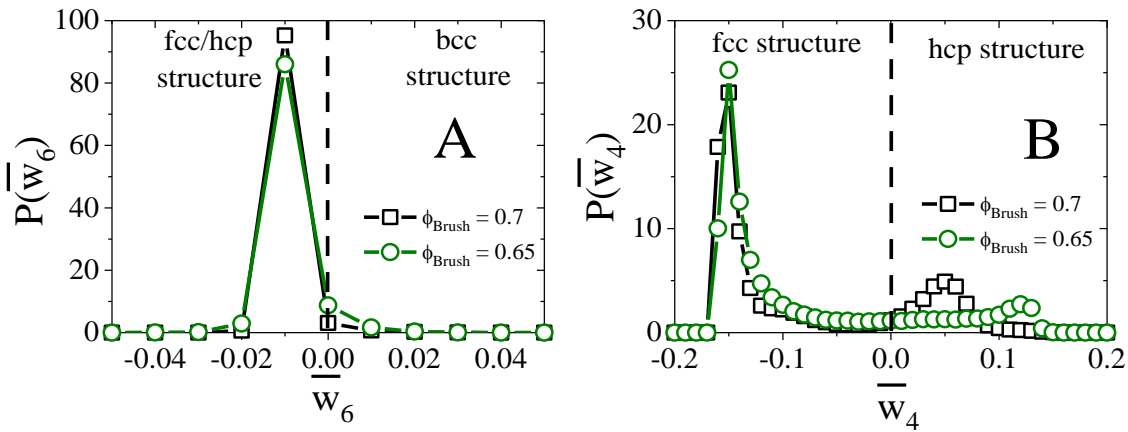

Figure 5. Rotationally invariant bond order parameter distribution A) $P(\overline{w}_6)$ and B) $P(\overline{w}_4)$ obtained with the effective brush potential.

**Dynamic properties**

Intermediate scattering functions at q = 0.025 nm$^{-1}$ are shown in Figure 6 as a function of the volume fraction, along with the corresponding relaxation time distributions with amplitude rescaled by the volume fraction. In the dilute non-interacting regime (ϕ = 0.076), or simply regime I, a single decay process is present. As the volume fraction increases to ϕ = 0.46, within the previously defined regime II, three distinct and physically meaningful diffusive (see Figure S6 in the SI) relaxation modes emerge. The fast process becomes faster with increasing volume fraction, undoubtedly describing the cooperative diffusion of the interpenetrated arms,[48,50,96] hence, polymeric in nature. The intermediate mode is nearly concentration-independent and attributed to the center of mass collective diffusion, as was also observed by Voudouris and coworkers in core-shell particles.[50] The slow mode, which slows-down with ϕ, is ascribed to PGNP long-time self-diffusion, detectable by light scattering due to the finite polydispersity of the particles.[48,50] This mode is present in both soft[40,45,50,96] and hard[97–100] colloidal systems.

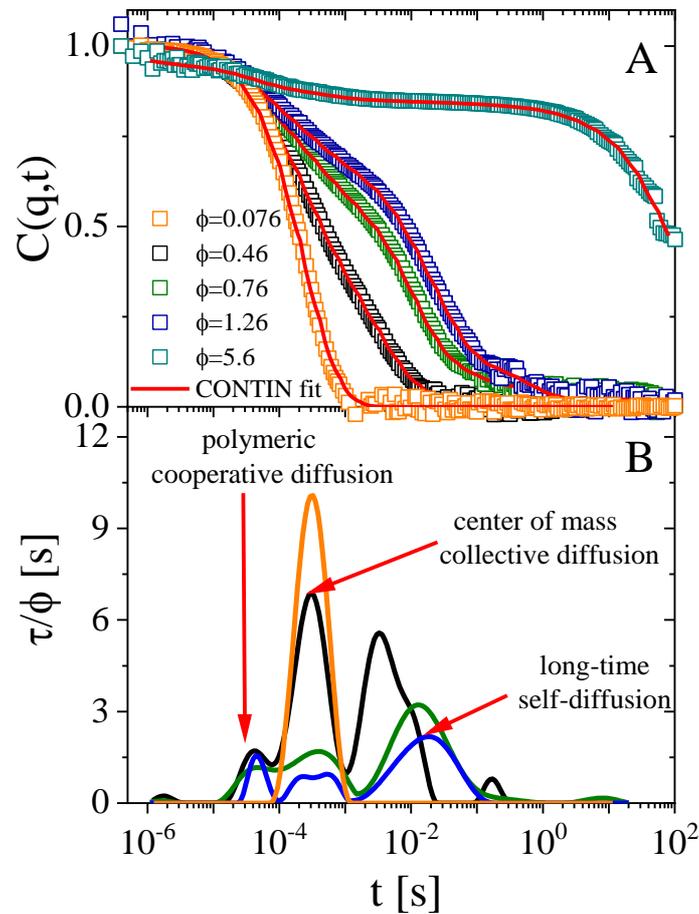

Figure 6. A) Intermediate scattering functions and B) relaxation time distribution divided by the volume fraction, at various volume fractions at a given q, 0.025 nm$^{-1}$, and at 25 °C. Transition from one single to three decay processes is shown. Data at ϕ = 5.6 (in the non-ergodic state) have not been analyzed because the time scale of the relaxation dynamics exceeds the resolution of the instrument. Red solid lines represent CONTIN fitting curves.

The concentration dependence of the light scattered intensity, I, normalized diffusion coefficient (by that in dilute regime), $D/D_0$, and the inverse of the relative viscosity, $\frac{1}{\eta_R} = \frac{\eta_{solvent}}{\eta_{solution}}$, is shown in Figure

7. In regime I, dilute non-interacting regime, the total intensity relates to the particle diffusion $D_0$ (see Figure 7A). The intensity of the polymeric cooperative mode constitutes a small contribution to the total intensity, and it decreases with $I \sim \phi^{-1.3}$ as reported in the literature.[48,53] Similarly, the intensity of the self-diffusion mode, decreases with increasing volume fraction, yet with a stronger dependence, $I \sim \phi^{-1.6}$, as also reported in the literature.[48,53] Finally, the center of mass collective motion exhibits a weak dependence of the intensity on volume fraction, as also observed in block copolymer micelles.[45]

In Figure 7B, until $\phi = 0.3$, in the dilute regime, the diffusion coefficient $D_0$ is constant and equal to 6.2 $10^{-12}$ m$^2$/s (see Figure S1 in the SI). At $\phi = 0.46$, in regime II, different modes appear. The fast mode speeds-up with concentration and follows the expected scaling for star and linear polymers in good solvent, $D \sim \phi^{0.77}$.[48,53–55] When chloroform is used as a solvent, the refractive indices of the core and the solvent are nearly matched, hence the contrast between the polymeric shell and the solvent is increased. The result is that only two modes are detectable (see Figures S7 and S8 in the SI): 1), the polymeric cooperative diffusion, which was further confirmed, and 2) the center of mass collective structural mode (see black symbols in Figure 7B). The center of mass collective diffusion (blue symbols) is nearly concentration-independent, as was also observed in core-shell particles,[50] whereas the self-diffusion (red symbols) significantly decreases with the volume fraction, and follows the same trend exhibited by the relative viscosity (green symbols) with a slight deviation at $\phi = 1.26$, at the onset of regime III. This finding, which confirms earlier findings with other stable block copolymer micelles[27,77] validates the Stokes-Einstein-Sutherland relation which assumes that the product between the self-diffusion coefficient and the zero-shear viscosity is constant: $\eta D = k_BT/6\pi R_H$. The origin of a deviation across the arrested state was thoroughly investigated by Gupta *et al.* and attributed to the emergence of dynamic heterogeneities.[101] In the light of this, it can be also hypothesized that, at the onset of the regime III, as a consequence of particles ordering, some structural modes manifest, complicating further the dynamic response of such soft systems. The inverse relative viscosity and the long-time self-diffusion were fitted to the Krieger-Dougherty equation,[102–104]

$$\frac{1}{\eta_R} = \left(1 - \frac{\phi}{\phi_m}\right)^{[\eta]\phi_m} \qquad (7)$$

where $\phi_m$ is the maximum packing fraction and $[\eta]$ is the intrinsic viscosity equal to 2.5 for spheres.[102,104] The empirical fit yielded a value of $\phi_m$ equal to 0.82, still belonging to regime II. Such a value is smaller compared to that expected for arrested state (crystal) ($\phi > 1.26$). However, the fact that $\phi_m > 0.74$, the maximum possible packing for hard spheres, reflects mild brush interpenetration, therefore, mild softness. This represents an additional signature that the system investigated in this work exhibits strong features of rigid spherical particles. Softer colloids typically exhibit much larger $\phi_m$.[96] In this respect, a special note on the highest-$\phi$ self-diffusion data point of Figure 7B is in order. It clearly deviates from the rest and the Krieger-Dougherty line. A tentative interpretation of this behavior calls for osmotic compression and/or arms interpenetration, as discussed in the context of other stable polymeric micelles and soft particles.[41,96,105,106] In this regime, the brush potential is likely not applicable. However, clarifying these interesting issues goes beyond the scope of the present investigation.

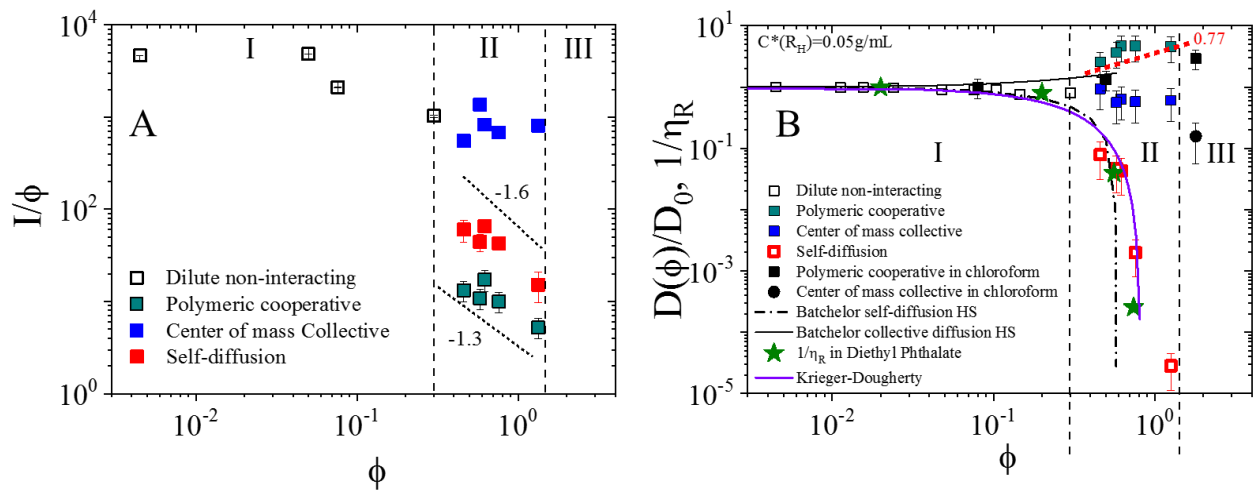

Figure 7. Volume fraction dependence of A) normalized intensity extrapolated at q→0 divided by the volume fraction, and B) normalized diffusion coefficients by that in the dilute non-interacting regime (regime I). Black solid symbols in panel B are measured in chloroform where the refractive index of the solvent nearly matches that of the core (see Table 3). The relative viscosity, green points in panel B, was measured in diethyl phthalate. Experiments were performed at 25 °C. Dotted black lines in panel A report theoretical predictions for the polymeric cooperative mode (I ~ $\phi^{-1.3}$) and particle self-diffusion (I ~ $\phi^{-1.6}$).[48] The dotted-dashed and solid black lines in panel B represent the Batchelor's predictions[107] for the long-time self-diffusion and center of mass collective diffusion for hard spheres (HS), respectively. The dotted red line in panel B refers to the theoretical prediction for the polymeric cooperative diffusion (D ~ $\phi^{0.77}$).[48] The solid purple line in panel B represents the fit of the inverse relative viscosity and long-time self-diffusion to the Krieger-Dougherty equation (see text). Vertical dashed lines mark the three regimes identified in Figure 2.

## Coarse-grained simulations

The static structure factor S(q) was calculated for both interaction potentials by solving the Ornstein-Zernike equation (OZ).[108] The latter involves an iteration procedure where different particle size values (hence, packing fractions) are tested in order to match the experimental pseudo-structure factor at $\phi = 0.5$, in the ergodic regime. Figure 8 shows the comparison between the experimental and the calculated static structure factor by using both star polymer and brush potential. The reported packing fractions and characteristic lengths used represent optimized values to match the experimental pseudo-structure factor (see SI for further details). Note that the obtained value of $\sigma_{SP}$ is very close to the measured hydrodynamic radius, $R_H = 62$ nm, while $\sigma_{Brush}$ resembles the hydrodynamic diameter, $R_H = 124$ nm, respectively. Remarkably, both interaction potentials provide quite a good agreement with the experimental results, with the star potential better capturing the amplitude of the first (main) peak of S(q), while being less efficient in capturing the position of the second-order peak. Nonetheless, given the experimental noise at large values of q, this does not represent a sufficient condition to discriminate the two interaction potentials. We will thus assess this point when comparing dynamical properties.

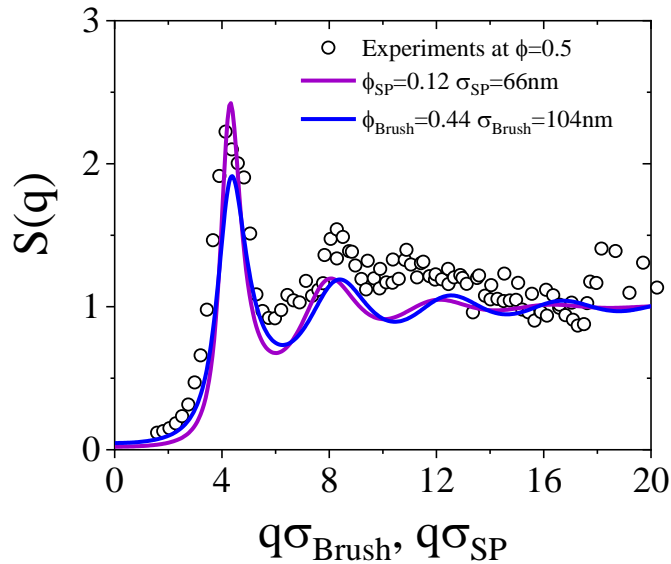

Figure 8. Experimental pseudo-structure factor at $\phi = 0.5$ along with the calculated ones by using both star polymer and brush potentials with the same number of arms, $f = N_{agg} = 212$. Values of particle size and packing fraction are optimized to match the experimental results (see text). In addition, for the brush potential $R_c/\sigma_{Brush} = 0.11$, $L/\sigma_{Brush} = 0.39$ $s = \sqrt{\frac{A}{N_{agg}}} = 3.65$.

Indeed, important differences arise when these interaction potentials are used to investigate particle dynamics. LD simulations were performed in order to obtain the concentration relaxation functions and compare them with the experimental intermediate scattering functions. The concentration range focuses on regime II. Figures 9, 10 and 11 depict the correlation functions and associated relaxation time distributions obtained with both potentials at three volume fractions, $\phi = 0.46$, $\phi = 0.76$ and $\phi = 1.26$ (close to the onset of regime III). As the experimental intermediate scattering functions were obtained at various scattering angles between 30° and 150°, the angle of 90° represents an intermediate angle or scattering wavevector which was chosen for the comparison. Experiments and simulations are compared at the same length scale, or more specifically, at the same scattering wavevector (q = 0.025nm$^{-1}$ at 90 °). This translates in varying $\sigma_{SP}$ and $\sigma_{Brush}$ so that the product $q\sigma_{SP}$ or $q\sigma_{Brush}$ complies with the experimental value (see SI for further details). The packing fraction is also adjusted accordingly. The experimental time is normalized by an arbitrary factor (1.5x10$^{-4}$ in simulation units) to match the dimensionless time obtained in simulations. This value was kept constant for all volume fractions and scattering wavevectors, for reasons of consistency. A key finding of this work is the satisfactory agreement between experimental data and simulations using the brush potential throughout the entire concentration regime (Figs.9-11). This agreement not only includes the slow relaxation process, but also the fast relaxation mode, a rather challenging task due to the presence of both self and collective modes, as shown in previous work with frozen block copolymer micelles.[101] On the other hand, the star polymer potential is not fully able to capture the experimental behavior at high volume fractions. This is evident in particular in Figs. 10 and 11. We note that several attempts to match the experimental data were performed for the star potential by varying either $\sigma_{SP}$ or $\phi_{SP}$, without being able to faithfully describe the full behavior of the experimental curve, as shown in Figure S9 of the SI for $\phi = 1.26$. Indeed, by varying $\phi_{SP}$ and or $\sigma_{SP}$, either the slow mode or the fast mode of the experimental intermediate scattering function can be separately captured, but not both at the same time. Conversely, the brush potential, characterized by the tunable parameter L, the brush layer thickness, which can also account for the osmotic compressibility of the shell, provides a better agreement with the experimental behavior. We believe that the key factor is the qualitatively different approach to repulsion as the interparticle distance decreases (i.e., different softness), which is depicted in Figure 1. Indeed, the star potential overestimates the softness of the PGNPs, resulting in a much smoother repulsion at particle contact. The presence of a highly cross-linked core shortens the range of repulsion between particles, making them harder compared to stars (see $R_g/R_H$ ratio in Table 2). In other words, one may consider the ratio between the total particle radius and the core radius as a parameter to predict the length of the particle interactions range. Below a certain value of this ratio, the star polymer potential does not seem to capture the behavior of a soft spherical system, while the brush potential is found to be valid. Although, this aspect deserves more investigation, literature works[8,29–31] combined with the present case, suggest as a threshold value a size ratio in the range 7-10 (present case ~ 5).

Relaxation time distributions were also computed for the simulated correlation functions and are reported along with the experimental ones in Figures 9, 10 and 11. The same CONTIN analysis as the experimental correlation functions was used for a fair comparison. The self-diffusion is reasonably well-captured by the simulated relaxation time distributions for all volume fractions. Surprisingly, the two fast processes seem to be present also in the simulated data at $\phi = 1.26$. However, it would be unreasonable to consider them separately and associate the fastest mode to the cooperative diffusion of arms with this level of coarse-graining. To this end, it seems more appropriate, and actually more consistent with the simulation model, to consider only two relaxation processes, the self-diffusion and the collective diffusion. The presence of a broad (at $\phi = 0.46$ and $\phi = 0.76$) and double-peak (at $\phi = 1.26$) in the faster mode is due to polydispersity. In fact, when the same simulations are performed without polydispersity, only one fast narrow relaxation process is observed (see Figure S10 of the SI).

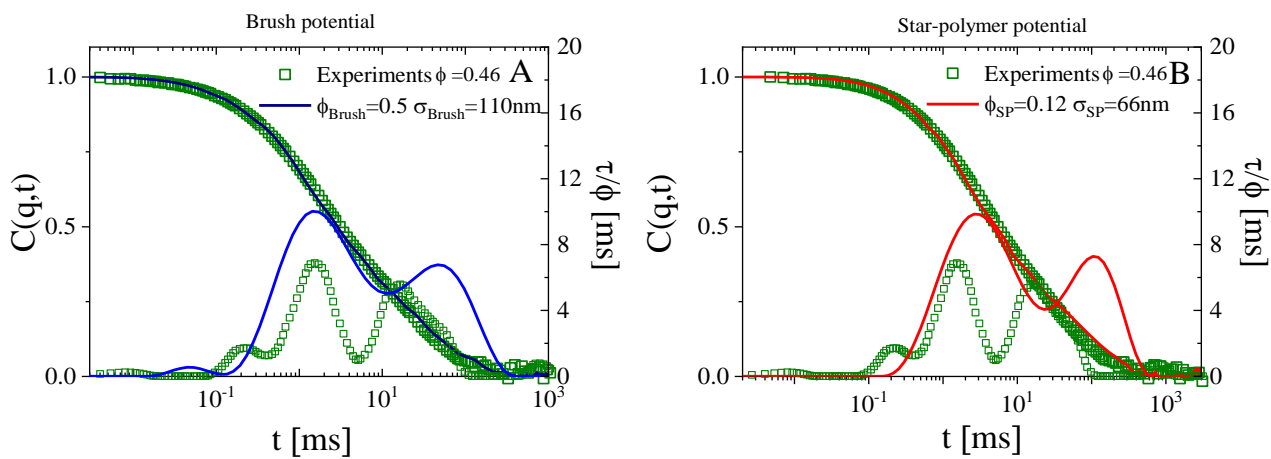

Figure 9. Experimental (symbols) intermediate scattering and numerical (lines) concentration relaxation functions along with the relaxation time distribution (divided by the volume fraction) for A) brush potential and B) star polymer potential. The experimental volume fraction is $\phi = 0.46$. The relaxation time distribution is obtained by CONTIN analysis. Both the experimental time and the relaxation time distribution are normalized by an arbitrary factor to match the dimensionless quantities obtained in simulations.

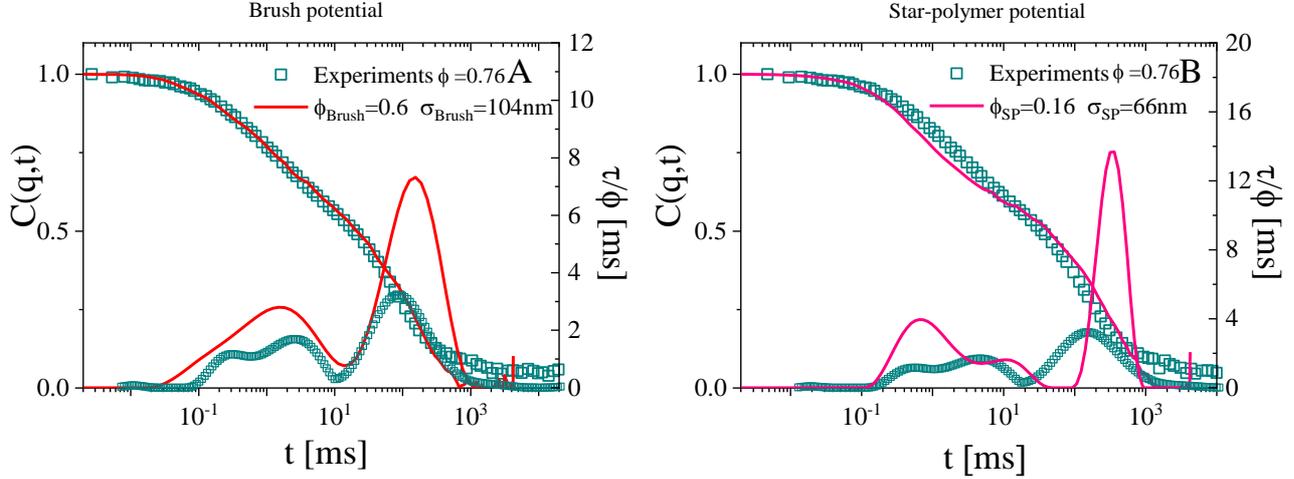

Figure 10. Experimental (symbols) intermediate scattering and numerical (lines) concentration relaxation functions along with the relaxation time distribution (divided by the volume fraction) for A) brush potential and B) star-polymer potential. The experimental volume fraction is $\phi = 0.76$. The relaxation time distribution is obtained by CONTIN analysis. Both the experimental time and the relaxation time distribution are normalized by an arbitrary factor to match the dimensionless quantities obtained in simulations.

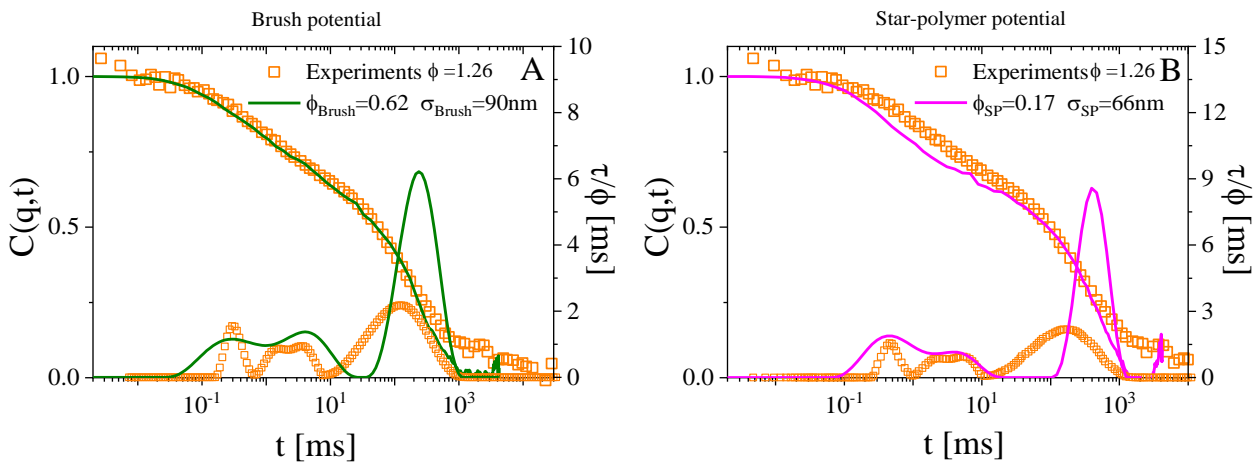

Figure 11. Experimental (symbols) intermediate scattering and numerical (lines) concentration relaxation functions along with the relaxation time distribution (divided by the volume fraction) for A) brush potential and B) star-polymer potential. The experimental volume fraction is $\phi = 1.26$. The relaxation time distribution is obtained by CONTIN analysis and subsequently divided by the volume

fraction. Both the experimental time and the relaxation time distribution are normalized by an arbitrary factor to match the dimensionless quantities obtained in simulations.

## Osmotic Compressibility

Star polymers are known to undergo osmotic de-swelling in a crowded environment,[83,109,110] an effect that can be pronounced depending on the functionality of the stars. In fact, for low to intermediate values of functionality (f < ~100) the osmotic de-swelling, quantified with the size reduction, resembles that of linear polymer chain solutions with $R(C) \sim C^{-1/8}$, with R being the star radius and C the concentration.[83] Star polymers with higher functionality (f~362) are more efficient osmotic compressors than linear chains since the osmotic pressure increases with functionality.[18,111] Osmotic pressure in high-functionality stars can also favor crystalline order.[1,28,112,113] In the present case, the shrinkage of the particles was estimated through the reduction of $\sigma_{Brush}$ needed to obtain good agreement between the simulated and experimental correlation functions when using the brush potential. Figure 12 depicts the shrinkage of the soft particles as a function of the volume fraction, using present data and star data from the literature.[18,29] Although only few data points are available, it is clear that the dependence of the particle size on the volume fraction ($\phi$) is much stronger for PGNPs compared to low functionality stars, and very similar to high-functionality stars, with a slope around -0.2.

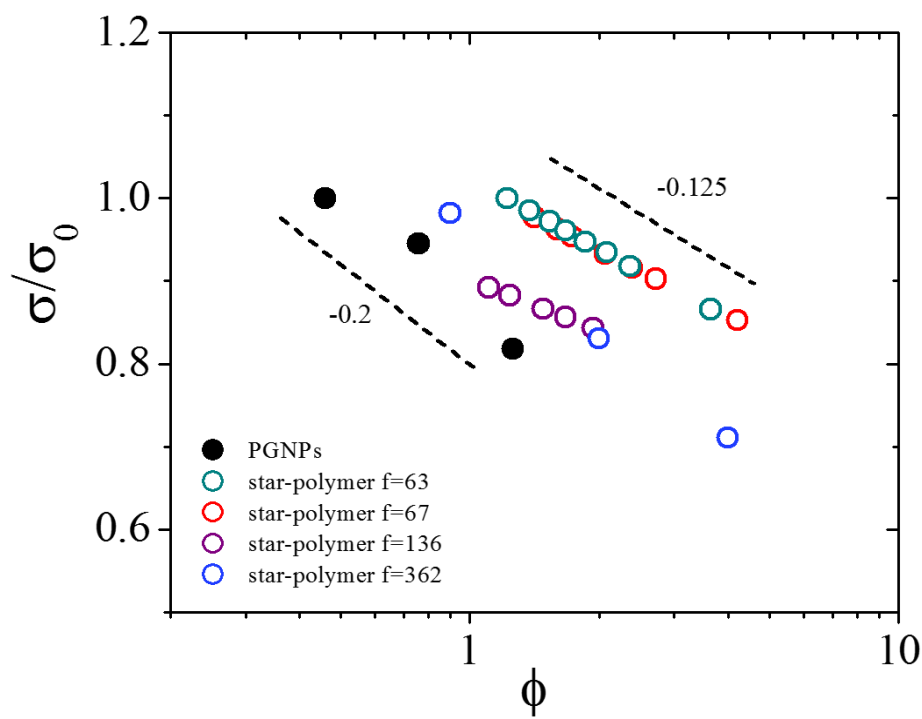

Figure 12. Normalized particle size by that at $\phi \to 0$ (no particle de-swelling) against volume fraction for PGNPs (solid circles) and star polymers taken from literature.[18,29,78] Dashed black lines represent the -1/8 slope predicted for linear polymer chains and low-functionality star polymers in solution, and -0.2 the slope found for high-functionality star polymers and PGNPs.

## Conclusions

A versatile, stable spherical model PGNP system with features encompassing core-shell particles and star polymers was investigated by means of experimental and simulation techniques with the goal of developing an experimental-simulations toolbox to describe the static and dynamic properties of such tunable colloidal systems over a large range of volume fractions from dilute to the non-ergodicity transition. In this particular case with negligible polydispersity and precisely controlled structure, the latter case is associated with crystallization of the PGNPs, marking a departure from the behavior of monodisperse multiarm star polymers which typically vitrify under similar conditions. This study allows testing the limits of the star-polymer interaction potential and obtain a better understanding on the dynamics of soft spherical colloids. The ergodic-to-non-ergodic transition upon increasing concentration was probed by light scattering and rheology. Neutron scattering experiments and simulations revealed that the non-ergodic state has features of a fcc crystalline structure. This is also predicted by simple brush conformation arguments, according to which, when the tethered chains belong to the semi-dilute polymer brush regime, the probability of having an fcc structure becomes higher than a hcp because the excluded volume interactions enhance the energetic and entropic differences between fcc and hcp phases.

Despite the relatively long-ranged repulsive interactions of the PGNPS compared to hard spheres, the star potential was not particularly effective in capturing their dynamics. The reason is that particle interactions of the investigated systems are not as long-ranged as those of star polymers. Combining literature works with the present one, suggests that when the ratio between the particle size and the core size is below 7-10, the star polymer potential cannot capture the dynamics of spherical soft colloids. Consequently, a clear need for an alternative interaction potential arose.

To this end, we adapted an existing brush interaction potential that captures reasonably well both the static and dynamic properties of the PGNPs at various concentrations at a fixed value of the scattering wavevector (q = 0.025 nm$^{-1}$ at 90°). Further investigations of this system and similar ones with different characteristic length scales, combined with targeted experimental and numerical studies on crystallization, will provide the possibility to have finesse the present results and come-up with a robust effective potential that describes a large class of core-shell particles, allowing the access to new insights on soft spherical nanoparticles.

We also showed that, particle shrinkage due to osmotic pressure was found to have a stronger dependence on volume fraction compared to simple linear chains or star polymers with relatively low functionality and long arms. For the PGNPs treated in this work, the dependence of the particle shrinkage on the volume fraction is about $\phi^{-1/4}$.

This work sets the foundation for a better understanding of the link between soft particle interactions and dynamics,[114,115] paving the way for a more raffinate design of soft hairy nanoparticles with desired range of static and dynamic properties.

## Supplementary Material

Static and dynamic light scattering in the dilute and semidilute regime in toluene and chloroform; comparison between simulations and experiments in terms of correlation functions by using the star polymer potential; effect of polydispersity in the simulated correlation functions.

## Acknowledgements

This research was supported by the EU (European Training Network COLLDENSE (H2020-MCSA-ITN-2014, grant number 642774 and Horizon2020-INFRAIA-2016-1, EUSMI grant no. 731019) and the National Natural Science Foundation of China (Grant No. 21674122). Helpful discussions with A. N. Semenov are gratefully acknowledged. Emanuela Zaccarelli and Jose Ruiz-Franco acknowledge support from the European Research Council (ERC Consolidator Grant 681597, MIMIC) and from Sapienza University of Rome through the SAPIExcellence program. Daniele Parisi acknowledges Dr. Fabien Dutertre for help with preliminary light scattering experiments.

## Data Availability Statement

The data that support the findings of this study are available from the corresponding author upon reasonable request.

# Supplementary Material

# Static and Dynamic Properties of Block-Copolymer Based Grafted Nanoparticles Across the Non-Ergodicity Transition

Daniele Parisi,[1] José Ruiz-Franco,[2] Yingbo Ruan,[3,4] Chen Yiang Liu,[3,4] Benoit Loppinet,[1] Emanuela Zaccarelli,[2,5] and Dimitris Vlassopoulos[1,6]

[1] Institute of Electronic Structure & Laser, FORTH, Heraklion 71110, Crete, Greece

[2] Dipartimento di Fisica, Sapienza Università di Roma, P.le A. Moro 5, 00185 Roma, Italy

[3] Beijing National Laboratory for Molecular Sciences, CAS Key Laboratory of Engineering Plastics, Institute of Chemistry, The Chinese Academy of Sciences, Beijing 100190, China

[4] University of Chinese Academy of Sciences, Beijing 100049, China

[5] CNR Institute of Complex Systems (ISC), Uos Sapienza, Roma, Italy

[6] Department of Materials Science & Technology, University of Crete, Heraklion 71003, Crete, Greece

## Table of Contents

# Light Scattering in dilute regime

Dynamic and static light scattering experiments in the dilute regime yielded the estimation of the diffusion coefficient, hydrodynamic radius and radius of gyration. Typical results are shown in Figure S1. Note that extrapolations at concentration and scattering wavevector to zero are needed in order to quantify the above-mentioned quantities.

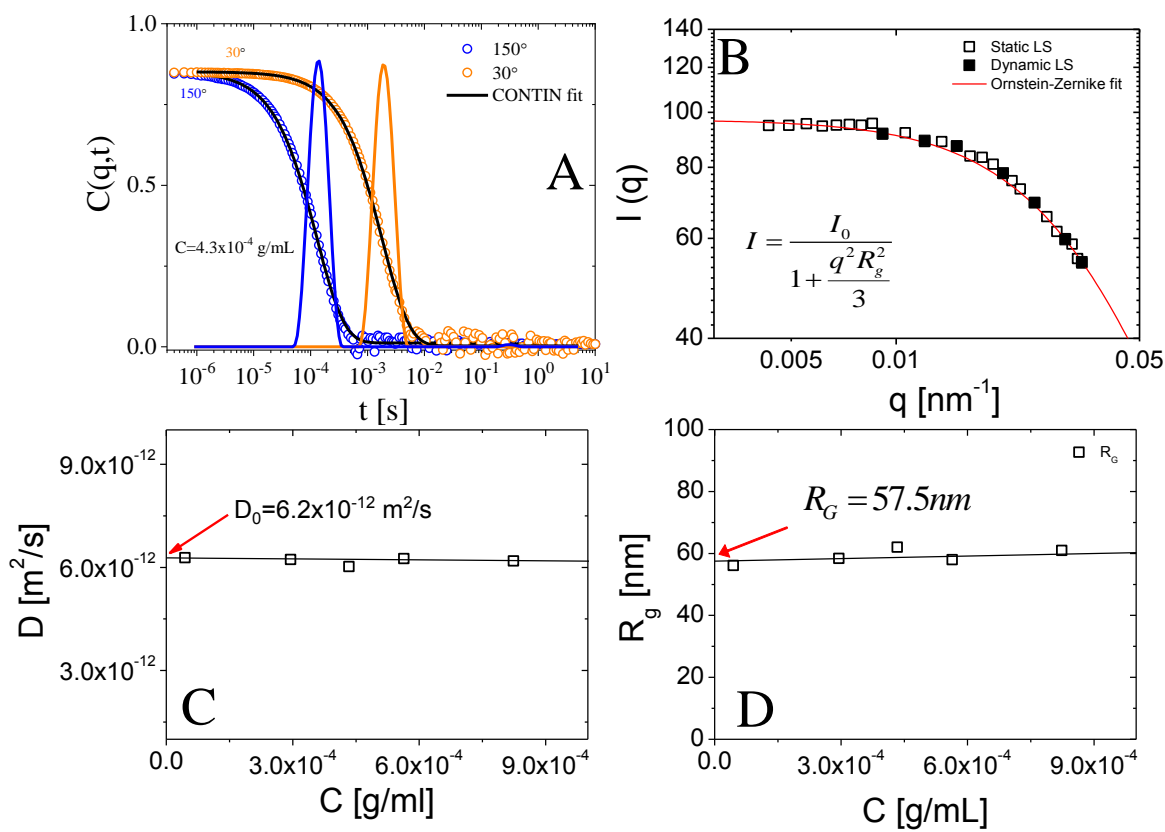

Figure S1. A) Intermediate scattering functions along with the relaxation time distribution at two different scattering angles, 30° and 150°, at C = 4.3 $10^{-4}$ g/ml. B) Normalized scattering intensity (to the solvent value) against the scattering wavevector q at the same concentration as in panel A. Intensities from dynamic light scattering measurements are also reported (solid black squares). Red solid line represents Guinier[1] fit to extract the radius of gyration $R_g$ as extrapolation to q = 0. C) Concentration dependence of the extracted diffusion coefficient. D) Concentration dependence of the extracted radius of gyration. Red arrows indicate values extrapolated at C = 0. Experiments were performed at 25 °C.

The form factor was obtained experimentally by means of Small-Angle Neutron Scattering measurements. An example is depicted in Figure S2 where the scattering intensity is fitted to the Perdersen and Gersterberg form factor for spherical block copolymer micelles in good solvency conditions, without adjustable parameters.[2] The radius of the core was measured through imaging analysis of TEM images in dry state (shown in the inset in Figure S2). The thickness was roughly estimated as the difference between the hydrodynamic radius and the radius of the core and found to be 48.5 nm. From the knowledge of the size of the core and the grafting density,[3] the aggregation number was estimated and amounted to 212.

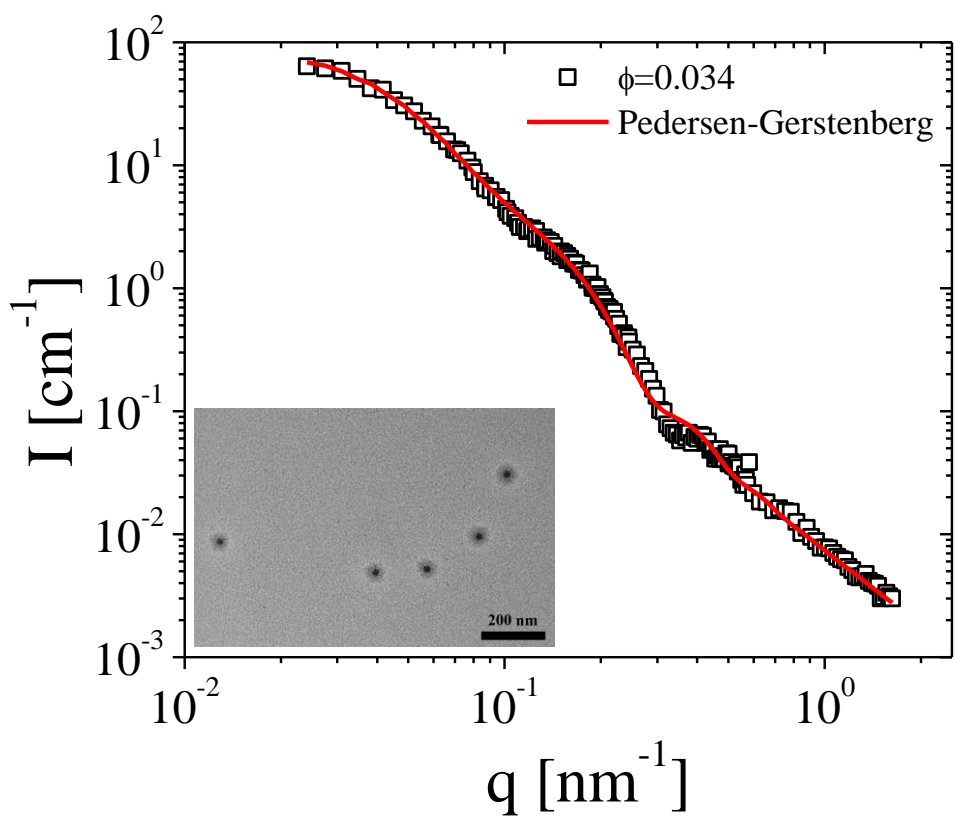

Figure S2. Normalized neutron scattering intensity against scattering wavevector q. Red line represents the fitting of the form factor with the Pedersen and Gerstenberg model.[2] No free parameters used in the fitting. Main input parameters: radius of the core (13.5nm), shell thickness (48.5 nm), aggregation number (212) and solvency conditions (good solvent). TEM image in dry state is shown in the inset. Experiments were performed at 25 °C.

Static light scattering measurements were performed in solutions at different concentrations well below c* in order to determine the solvency conditions by using the Zimm equation.[4] The equation already extrapolated to zero scattering wavevector (q→0) has the following form and is plotted against the concentration in Figure S3.

$$\frac{C}{R_\theta} = \frac{1}{K}\left(\frac{1}{Mw} + 2A_2C\right) \quad (1)$$

where C is the concentration in g/ml, K the optical contrast constant (dimensionless)[4] defined as

$$K = \frac{4\pi^2 n_{sol}^2 \left(\partial n/\partial c\right)^2}{\lambda^4 \mathcal{N}_A} \quad (2)$$

where $n_{sol}$ is the refractive index of the solvent (see Table 3), $\partial n/\partial c$ is the refractive index increment and $\lambda$ is the laser wavelength. $R_\theta$ represents the excess Rayleigh ratio[4] in cm$^{-1}$ defined as

$$R_\theta = \frac{I_\theta - I_{sol}}{I_{tol}} R_{tol} \left(\frac{n_{sol}}{n_{tol}}\right)^2 \quad (3)$$

with $I_\theta$ being the total scattered intensity at the scattering angle θ, $I_{sol}$ the scattering intensity of the solvent, $I_{tol}$ the scattering intensity of the standard solvent (toluene), $R_{tol}$ the Rayleigh ratio of the standard solvent (toluene), $n_{tol}$ the refractive index of the standard solvent (toluene) (see Table 3), and $A_2$ the second virial coefficient. Trivially, K can be found by extrapolating the experimental data to zero-concentration and the second virial coefficient through the slope (see Figure 2). From Equation 7 it was found that the refractive index increment is 0.12 ml/g, very close to that of pure PS in toluene at the same laser wavelength, 0.11 ml/g.[5] The second virial coefficient $A_2$ resulted positive, implying good solvency conditions, but its value 3.4 10$^{-5}$ mol ml g$^{-2}$ is almost one order of magnitude lower than that of pure PS chains in toluene at the same $M_w$, 1.1 10$^{-4}$ mol ml g$^{-2}$.[6] As expected, particle interactions are different from those between simple linear chains. This relates to the colloidal nature of the PGNPs.

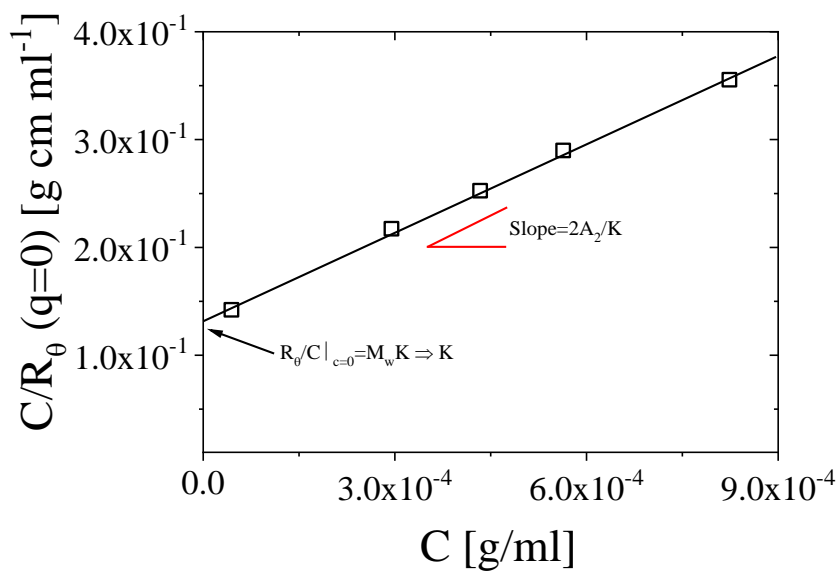

Figure S3. Concentration normalized by the excess Rayleigh ratio extrapolated at the scattering wavevector q tending to zero against concentration. The K parameter was extrapolated at C = 0, 2.5 $10^{-7}$ mL mol $g^{-2}$ $cm^{-1}$. Whereas the second virial coefficient $A_2$ from the slope, 3.4 $10^{-5}$ mol ml $g^{-2}$. The refractive index increment was found to be 0.12 ml/g, very close to that of pure PS in toluene at the same laser wavelength, 0.11 ml/g.[5]

## Dynamic light scattering in semidilute regime

In Figure S4A Intermediate scattering functions at different volume fractions but at the same scattering wavevector ($\theta = 90°$) are shown together with a double KWW fit. Shape exponents are displayed in Figure S4B. Whereas the slow mode at different concentrations exhibits a stretching exponent $\beta_1$ between 0.8 and 0.9 (i.e., not far from single exponential decay), the fast one has $\beta_2$ values around 0.5. Hence, treating the fast mode as a one single broad relaxation mode or two distinct modes represents a challenge. It is important to emphasize that a double stretched exponential function has no true physical meaning. Nevertheless, it revealed useful to identify relaxation modes. To shed light on this aspect, Figure S5 shows the analysis of an intermediate scattering function at a given angle and concentration. CONTIN[7] fit can yield either a broad fast relaxation time distribution covering almost three decades, or two nearly exponential fast relaxation modes. The three-mode analysis of the C(q,t) was considered appropriate to describe the dynamics of such soft nanoparticles, with the slow mode always treated as a single exponential process.

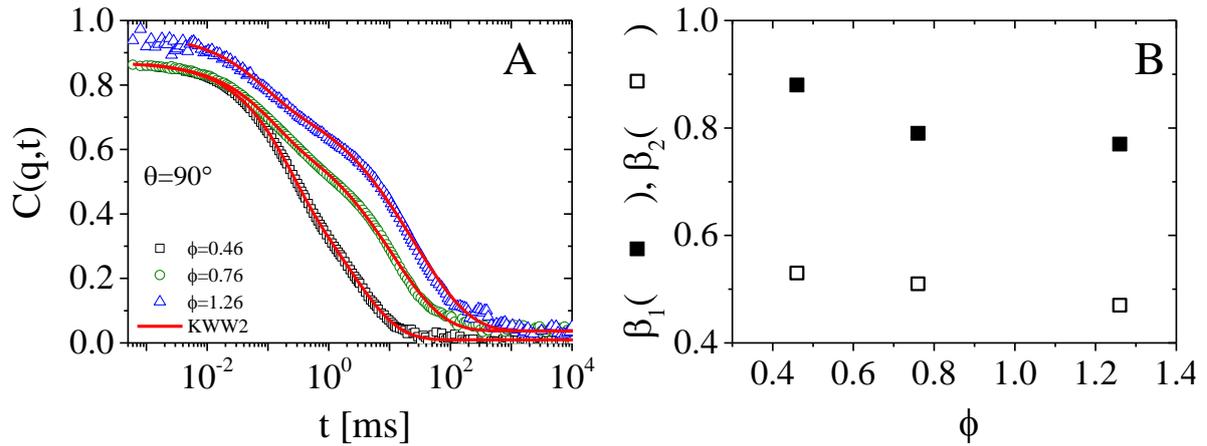

Figure S4. A) Intermediate scattering functions for three different volume fractions at the same scattering wavevector. Red lines represent fitting curves with a double KWW expression. Although the double stretched exponential function does not have any physical meaning, it is often used in light scattering to investigate the broadness of the relaxation time distributions. B) Fast (open squares) and slow (solid squares) stretch exponents as a function of the volume fraction. Values around 0.5 imply that the relaxation process is rather broad.

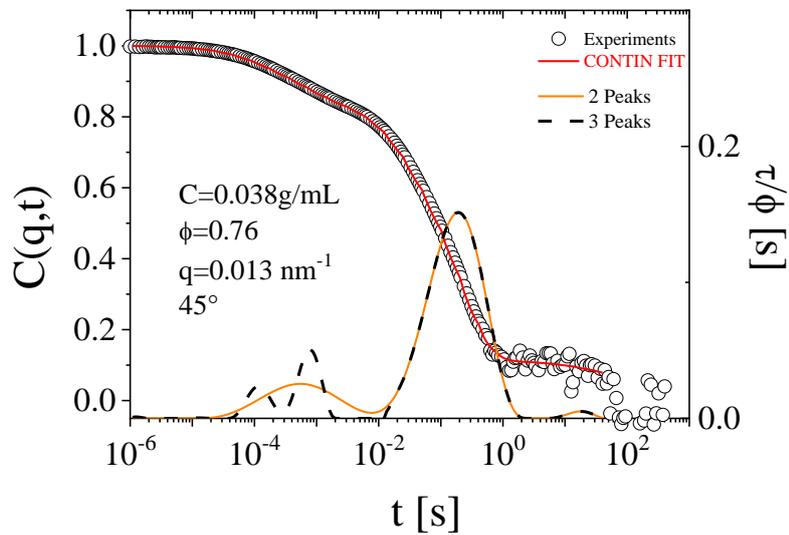

Figure S5. Intermediate scattering functions and relaxation time distribution (divided by the volume fraction) at a given volume fraction and scattering wavevector reported in the legend. One very broad fast relaxation time distribution is shown along with the two single exponential modes analysis. Different evidences suggest that two fast modes can be decoupled and assigned to two different relaxation processes (see text). Note that both analyses provide the same CONTIN fit, as only the relaxation time distribution has been changed.

In Figure S6, the q-dependence of the decay rate for the three modes detected in Figure 9 of the manuscript is shown for different volume fractions, $\phi = 0.46$, $\phi = 0.76$, $\phi = 1.26$. The error bars refer to the estimation of the relaxation times with the CONTIN analysis. Within measurement error, the three modes are diffusive, as expected for polymeric cooperative, center of mass collective and self-diffusion.[8,9]

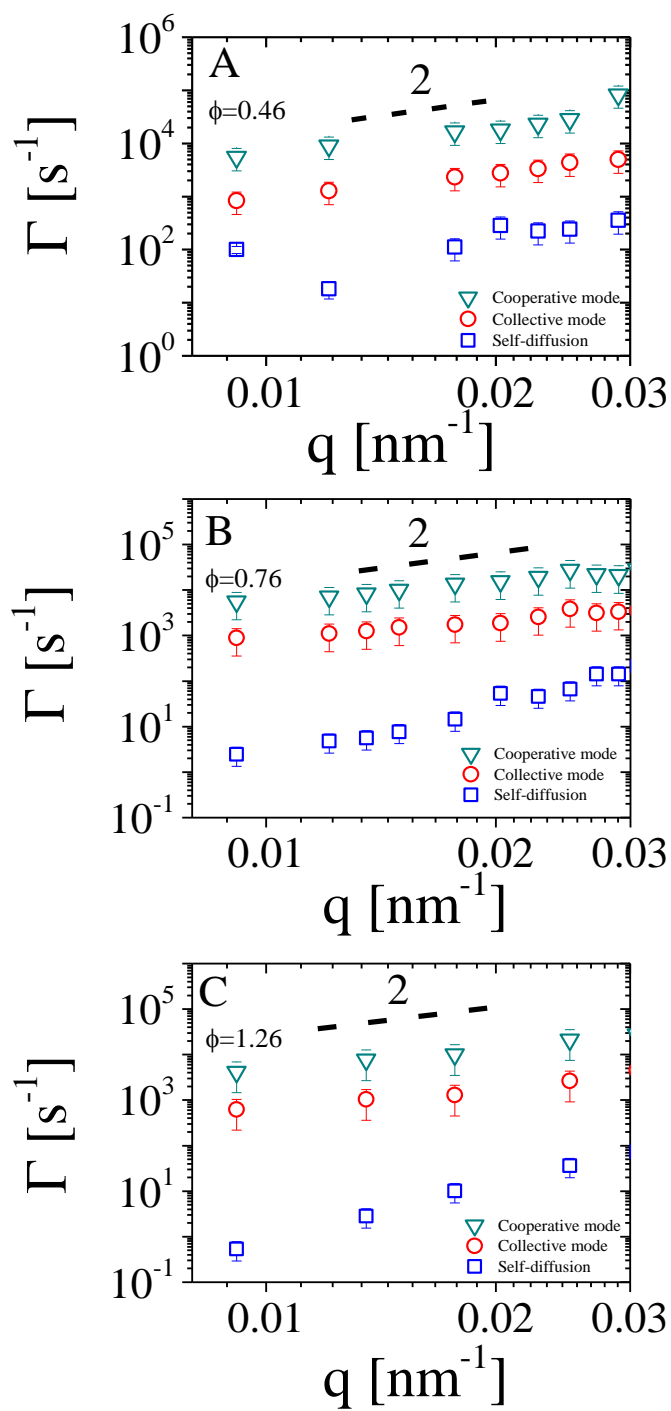

Figure S6. Decay rate Γ as a function of the scattering wavevector q at three different volume fractions: $\phi = 0.46$ (panel A), $\phi = 0.76$ (panel B) and $\phi = 1.26$ (panel C). Dashed black line indicates the slope of a diffusive process. Experiments were performed at 25 °C.

## Dynamic light scattering in chloroform

Figure S7 depicts the experimental intermediate scattering functions of PGNPs in chloroform at a fixed scattering wave vector and different concentrations. The rationale behind the use of chloroform is to obtain a nearly iso-refractive conditions between the core and the solvent and increase the shell-solvent contrast to make the polymeric cooperative mode more visible. The q-dependence of the normalized intensity is reported in Figure S8. Besides the fraction $\phi = 0.08$ where only one process is discerned and the intensity decreases with q, the two larger fractions $\phi = 0.5$ and $\phi = 1.84$ show an increase of the intensity with increasing the scattering wavevector.

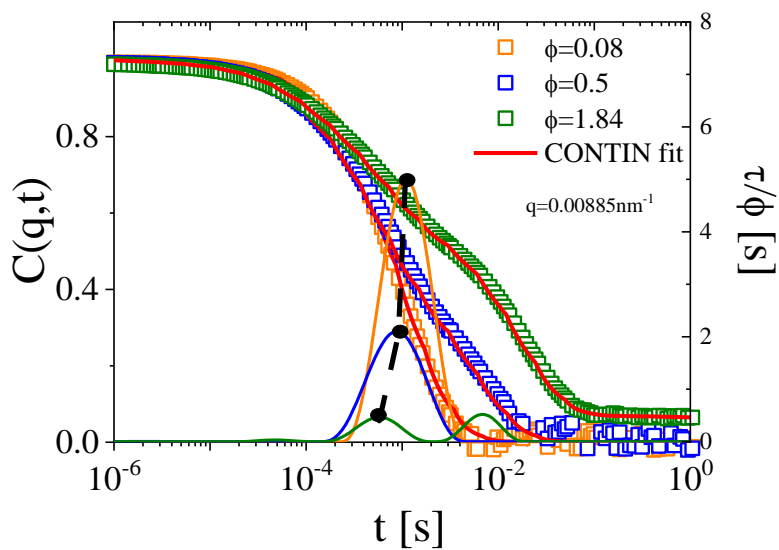

Figure S7. Intermediate scattering functions at various volume fractions at q=0.00885 nm$^{-1}$. In the right-hand y-axis the relaxation time distribution rescaled by the volume is plotted. Dashed black

lines and black dots guide the eye to the speed-up of the dynamics with volume fraction. Red lines represent CONTIN fitting curves. Experiments were performed at 25 °C.

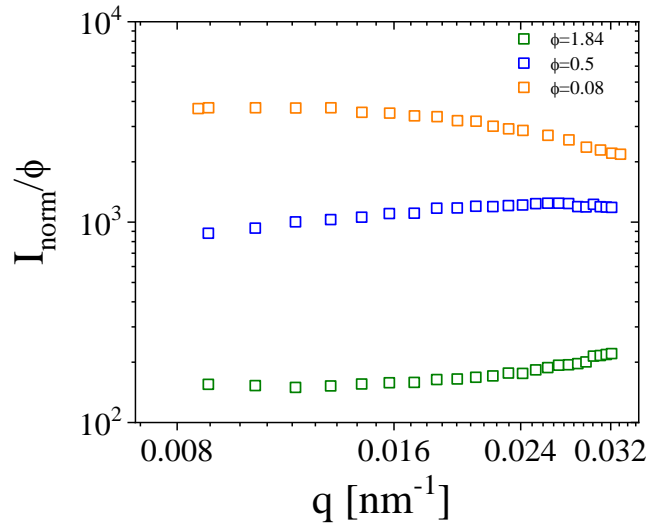

Figure S8. Total normalized intensity (by the solvent) divided by the volume fraction against the scattering wavevector at various volume fractions. Experiments were performed at 25 °C.

## Comparison between experiments and simulations

As reported in the main text, the comparison between simulations and experiments is conducted at the same scattering wavevector q. While for the experimental system this is fixed to q = 0.025 nm$^{-1}$ (90°), in the simulations the wavevector is constrained to be an integer multiple (say n) of the box mesh in Fourier space, i.e. $\left(\frac{\pi n}{L_{BOX}}\right)$ and it will be provided in units of the inverse of the characteristic length, $\sigma_{Brush}$ or $\sigma_{SP}$. To compare experiments and simulations, we thus fix N and vary the packing fraction via $L_{BOX}$. Then we select the values of n and length (in real units) which gives the closest value to the experimental wavevector.

In Figures 9 and 10 of the main text we showed that both potentials, with appropriate packing fraction, are able to capture the experimental correlation function below the overlap concentration. However, in Figure 11, the failure of the star potential was observed. One may wonder, whether or not a better agreement can be reached while further changing packing fraction and or particle size. Figure S9 shows relaxation functions obtained by simulations using the star polymer interaction potential at various packing fractions ($\phi_{SP}$) and selected particle sizes ($\sigma_{SP}$), along with the experimental results at $\phi$ = 1.26. The finding suggests that by varying $\sigma_{SP}$ and or $\phi_{SP}$, it is possible to capture the fast mode or the slow mode, but not both at the same time. On the contrary, the brush potential, well captures the experimental data by considering a small reduction of particle size. The latter can be taken into account in the brush potential just by changing the brush layer thickness, keeping constant the radius of the core, the grafting density and the aggregation number.

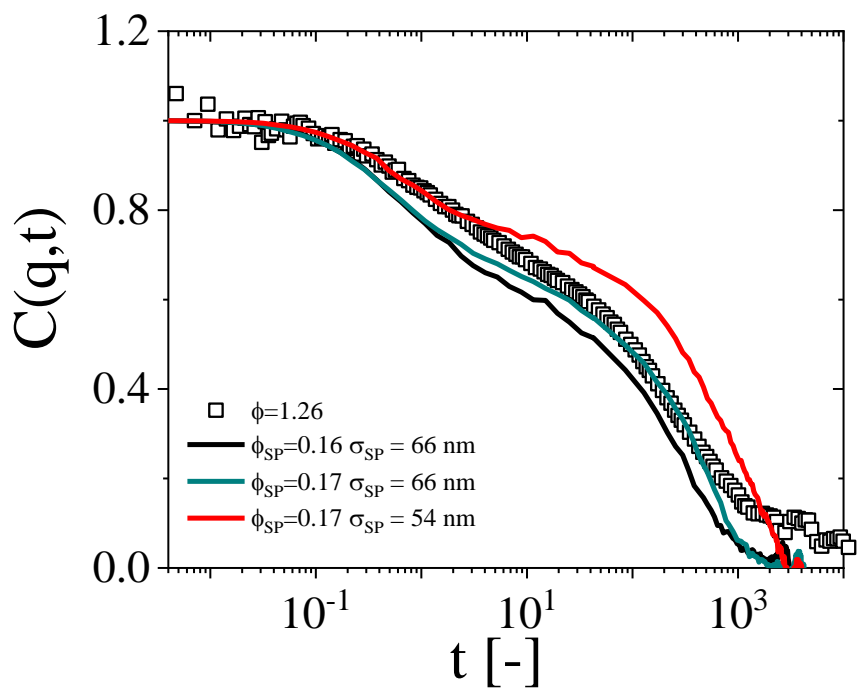

Figure S9. Comparison between experiments at $\phi = 1.26$ and simulations with star polymer potential at various packing fraction $\phi_{SP}$ and or particle size $\sigma_{SP}$, in terms of concentration relaxation function from simulations (lines) and intermediate scattering function from experiments (open symbols). No optimizing parameters were found. Experiments were performed at 25 °C. The experimental time is normalized by an arbitrary factor to match the dimensionless time obtained in simulations.

Simulations as well as experiments show three relaxation processes when the same CONTIN analysis is performed to the simulated concentration relaxation functions. The origin of a very fast process in simulations is totally different compared to that of experiments. Whereas the experimental very fast mode was assigned to the cooperative diffusion of the interpenetrating arms,[8] coarse-grained simulations do not detect such a mode as there is no distinct polymeric contribution taken into account. In fact, the presence of a fast mode in simulations is attributed to polydispersity as it is shown in Figure S10. Note that the polydispersity was introduced into the system in order to avoid crystallization and when this is reduced or removed the fastest process disappear and only the center of mass collective and self-diffusion are present.

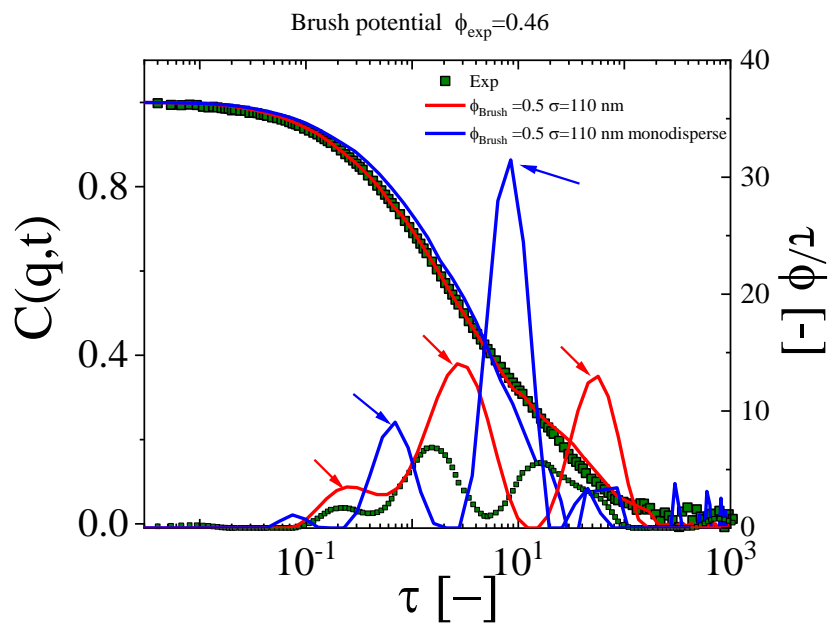

Figure S10. Concentration relaxation functions from simulations along with the relaxation time distribution divided by the volume fraction: polydispersity effect. The effective brush potential was used. When polydispersity is not considered, a narrow single fast process is shown (blue dashed lines). Arrows indicate the peaks in the relaxation time distribution.